\documentclass[a4paper,10pt]{article}

\usepackage{cite}

\usepackage{times}

\usepackage[utf8]{inputenc}
\usepackage[dvips]{graphicx}
\usepackage{amssymb}
\usepackage{latexsym}
\usepackage{amsmath}
\usepackage{lineno}
\usepackage{color}
\usepackage{xspace}
\usepackage{caption}

\newcommand{\ie}{\textit{i.e.}~}

\newcommand{\sO}{\textbf{O}\xspace}
\newcommand{\sF}{\textbf{F}\xspace}
\newcommand{\sC}{\textbf{C}\xspace}
\newcommand{\sD}{\textbf{D}\xspace}

\newcommand{\sFRC}{\textbf{FR-C}\xspace}
\newcommand{\sFRD}{\textbf{FR-D}\xspace}
\newcommand{\sSC}{\textbf{SC}\xspace}
\newcommand{\sSD}{\textbf{SD}\xspace}

\newcommand{\sFRO}{\textbf{FR-O}\xspace}
\newcommand{\sFRF}{\textbf{FR-F}\xspace}

\newcommand{\sSCO}{\textbf{SC-O}\xspace}
\newcommand{\sSDO}{\textbf{SD-O}\xspace}

\newcommand{\sB}{\textbf{B}\xspace}
\newcommand{\sBp}{\textbf{B$^+$}\xspace}
\newcommand{\sR}{\textbf{R}\xspace}
\newcommand{\sRp}{\textbf{R$^+$}\xspace}
\newcommand{\sS}{\textbf{S}\xspace}
\newcommand{\sSp}{\textbf{S$^+$}\xspace}
\newcommand{\sSR}{\textbf{S+R}\xspace}

\newcommand{\SM}{Supplementary Material\xspace}
\newcommand{\EM}{Extended Methods\xspace}

\title{Signalling boosts the evolution of cooperation in repeated group interactions}

\author{Luis A. Martinez-Vaquero,$^{1,2\ast}$ Francisco C. Santos,$^{3}$ Vito Trianni$^{1}$\\
	\\
	\normalsize{$^{1}$Institute of Cognitive Sciences and Technologies
		National Research Council of Italy,}\\
	\normalsize{ via San Martino della Battaglia 44, 00185 Rome, Italy}\\
	\normalsize{$^{2}$\textit{Current address}: International Institute for Sustainability,}\\
	\normalsize{estrada Dona Castorina 124, Rio de Janeiro, Brazil }\\
	\normalsize{$^{3}$INESC-ID Lisboa and Instituto Superior T\'ecnico, Universidade de Lisboa, Lisboa, Portugal}\\	
	\\
	\normalsize{$^\ast$To whom correspondence should be addressed; E-mail: l.martinez.vaquero@gmail.com.}
}

\date{}

\begin{document}
	
\maketitle

\section*{Abstract}
Many biological and social systems show significant levels of
collective action. Several cooperation mechanisms have been proposed,
yet they have been mostly studied independently. Among these, direct
reciprocity supports cooperation on the basis of repeated interactions
among individuals. Signals and quorum dynamics may also drive
cooperation. Here, we resort to an evolutionary game theoretical model to jointly analyse these two mechanisms  and
 study the conditions in which evolution selects for direct
reciprocity, signalling, or their combination.
We show that signalling alone leads to higher levels of
cooperation than when combined with reciprocity, while offering additional
  robustness against errors. Specifically, successful strategies in
the realm of direct reciprocity are often not selected in the presence
of signalling, and memory of past interactions is only exploited
opportunistically in case of earlier coordination
failure. Differently, signalling always evolves, even when costly. In
the light of these results, it may be easier to understand why direct
reciprocity has been observed only in a limited number of cases among
non-humans, whereas signalling is widespread at all levels of
complexity.

\section*{Keywords}
		Cooperation, evolutionary dynamics, game theory, signalling, reciprocity, public good games

	\section{Introduction}
	
	Cooperation and collective action problems are pervasive at all levels
	of biological complexity, from bacteria~\cite{miller2001quorum,
		Cornforth:2012fa,Nadell:2016bl, skyrms2010signals} to the most
	complex eusocial animals~\cite{Queller:2000cv} and humans
	\cite{Smith:2010jn}. In many situations, performing a task which is
	beneficial to an entire group demands a coordinated action of some
	kind, whereby individuals need to agree upon the action they will
	perform to survive or fulfil tasks that provide benefits to the group
	\cite{skyrms:2003, skyrms2010signals, souza2009evolution,kurokawa2009emergence,
		pacheco2009evolutionary, pacheco2015co,Pena:2015km,Gavrilets:is}.  Generally
	speaking, the individual costs associated to collective action, that
	is, costs incurred by the single to the benefit of the group, may
	bring forth coordination problems and social dilemmas. And with
	them, the shadow of free-riding and exploitation, and its direct
	consequence, the tragedy of the commons~\cite{hardin1968tragedy,
		rand2013human, Gardner:2008cn}.
	
	The mechanisms and behaviours that support cooperation and
	collective action vary widely, from the emergence of
	leaders~\cite{King:2009bx,Johnstone:2011cz} to the evolution of
	communication and language~\cite{Hauser:2002uy}, and are bound to the
	development of cognitive abilities that suitably balance the costs and
	benefits related to the collective action outcome. Coordination often
	requires information transfer through interactions among individuals,
	which may take many different forms, from indirect stigmergic
	processes to the direct exploitation of cues or signals
	\cite{wilson-sociobiology,maynard-smith:1995,stigmergy,
		skyrms2010signals}. According to Maynard Smith and Harper 
	\cite{maynard03:_animal}, a ``signal'' is defined as ``an act or
	structure that alters the behaviour of another organism, which evolved
	because the receiver’s response has also evolved''. Signals are known
	to be used as helpers of cooperation, despite they may entail a
	significant cost to the
	signaller~\cite{gintis2001costly,skyrms2010signals}.
	Other explanations have been proposed to account for the pervasiveness
	of cooperation in nature~\cite{nowak:2006b, rand2013human}. Some are
	based on memory, and direct reciprocity is probably the mechanism that
	received more attention~\cite{trivers:1971,axelrod:1984}. Direct
	reciprocity assumes that, when two players meet in a strategic
	interaction, they may prefer to cooperate if there is a high chance
	that they meet again. The same principle can be extended to N-person
	interactions. While the range of possible
	conditional strategies can be much larger than in 
        the 2-person case,
	reciprocity can still efficiently discourage individuals from
	free-riding in N-person collective~\cite{van2012emergence, pinheiro2014evolution,
		hilbe2017memory, martinez:2012,kurokawa2018evolution}. 
Direct reciprocity has also been observed in non-human animals~\cite{milinski1987tit}. However, these cases are rare when compared with humans~\cite{hammerstein:2003b,clutton2009cooperation},
	possibly due to the
	complex cognitive abilities required to reciprocate, ecological
	constraints or evolutionary bootstrapping problems~\cite{dunbar1998grooming,Hauser:2009ju,WHITLOCK:2007kz,Andre:2014kda,taborsky2016correlated,moreira2013individual,santos2018social}. This suggests that, in the presence of
	other promoters of cooperation (e.g., signalling~\cite{skyrms2010signals}), natural selection
	may favour simpler setups than direct reciprocity. On the other hand,
	repeated interactions may be prone to foster more efficient and honest
	signalling~\cite{rich2016honesty}, creating a valuable synergy for
	the emergence of cooperation.

	Evolutionary game theory (EGT) has successfully been used to study the
	emergence of cooperation both in the presence of direct reciprocity
	\cite{nowak:2006b, van2012emergence, pinheiro2014evolution,
		hilbe2018partners} and signalling~\cite{skyrms2010signals,
		pacheco2015co}, but the interplay between the two has
	received little attention~\cite{lotem2003reciprocity}.
	In particular, the evolutionary dynamics emerging from signalling
	in $N$-person dilemmas have been thoroughly analysed
	~\cite{pacheco2015co}, however without taking into account the
	possibility of reciprocation.
	Here we include this possibility, significantly enlarging the strategy
	space, investigating to which extent cooperation benefits from the
	interplay of more than one mechanism known to independently foster
	it. To do so, we propose a novel evolutionary game theoretical model
	in which we combine  signalling  with the possibility
	to reciprocate previous actions. We analyse under which conditions
	one, the other or a combination of them is evolutionary
	advantageous, assuming well-mixed populations and pairwise imitation evolutionary dynamics~\cite{traulsen:2006a}. By doing so, we are able to analyse the complex ecology
	resulting from the relation between reciprocity and signalling, and
	their co-evolution.
	
	Our results suggest that these two mechanisms, when combined, do not
	act synergistically to foster higher levels of cooperation. We show
	that thriving reciprocity strategies are not selected in the
	presence of signalling and that individuals do better by resorting to
	costly signalling, neglecting information from past encounters. We
	find that the combination of signalling and memory-based strategies is
	undermined by the emergence of a new class of opportunistic strategies
	that contribute to the common good  when the previous attempt
	of coordination failed and conversely free-ride after a successful
		cooperation event, overall leading to lower levels of cooperation.

	\section{Model}
	\label{sec:model}

	We consider a repeated $N$-person coordination problem, whereby a
		new round is played with probability $\omega$. A sequence of rounds
		can be played against either of two opposed ecological
	conditions~\cite{pacheco2015co}:
	a public goods game $G$ with probability $\lambda$---where
	individuals get a benefit proportional to the number of cooperators if
	a minimum number of them $M$ has been reached---and a
	non-public-goods-game $\tilde{G}$ with probability
	$(1-\lambda)$---where all the individuals obtain the same payoff
	independently of their actions. These ecological conditions can be
	interpreted as starvation and abundance states~\cite{pacheco2015co},
	where the former corresponds to $G$ (enough individuals should
	cooperate in order to get any benefit) and the latter to $\tilde{G}$
	(the ecological condition does not entail a public goods problem, so
	that decisions do not have any influence and all individuals obtain
	the same payoff). This setting allows for the assessment of the
	evolutionary dynamics under environmental variation, here associated
	with adverse or favourable conditions for a collective of
	individuals. Equivalently, we can imagine a value-sensitive decision
	making problem~\cite{Pais:2013ek} in which $G$ represents a highly
	valuable public good for which individuals need to spend themselves,
	and $\tilde{G}$ represents instead an unprofitable condition for which
	individuals should better avoid investing energies.
	Notwithstanding the interpretation of the ecological conditions, we
	consider that individuals interact through a repeated non-linear
	public goods problem~\cite{pacheco2015co,
		pacheco2009evolutionary}, where cooperating individuals incur in a cost
	$c$ while defectors pay no such cost. In each round, some form of coordination
	among individuals is needed to achieve a collective benefit. If a
	minimum number $M$ of cooperators is reached in a round, all
	individuals receive a benefit $b=rcN_C/N$, where $N_C\ge M$ is the
	number of cooperators and $N$ the total number of individuals in the
	group. The multiplication factor is $r>1$ when in $G$, and
	$r=0$ when in $\tilde{G}$. Moreover, when signalling is considered,
	individuals that signal incur in an additional cost $c_s$ (see \EM in \SM
	for details).
	
	In our model, individuals adopt strategies that are conditional on the past actions
	and on signals of the group mates,
	encoding the conditions for which individuals act in one or the other
	way~\cite{boyd:1988,kurokawa2010generous, santos2011co,pacheco2015co}.
	We consider the possibility of signalling as a function of the
	ecological condition, with two binary choices ($s_{G}$ and
	$s_{\tilde{G}}$) indicating if an individual signals (1) or not (0)
	respectively in state $G$ or $\tilde{G}$. Similarly to
	\cite{pacheco2015co}, we do not give any a priori meaning to such
	signals, and let the corresponding actions evolve freely in response
	to the attainment of a signalling quorum $Q$: individuals can
	cooperate (1) or defect (0) as a function of the number of signalling
	individuals, whether this exceeds the quorum ($a_{Q}$) or not
	($a_{\tilde{Q}}$). Similarly, the memory of past interactions can be
	exploited at each round: individuals can cooperate or defect as a
	function of whether the threshold $M$ of cooperators in the previous
	round was reached ($a_{M}$) or not ($a_{\tilde{M}}$).
	Execution errors are introduced for both signals and
          actions as a small probability $\epsilon$ that the individual chooses the opposite option that she intends to.
	When both
	signalling and reciprocity are present, individual strategies are
	represented by six bits:
	$[s_{G}, s_{\tilde{G}}; a_{\tilde{Q},M}, a_{\tilde{Q},\tilde{M}},
	a_{Q,M}, a_{Q,\tilde{M}}]$,
	where the first two bits indicate the signalling strategy, and the
	last four bits indicate the actions corresponding to the attainment of
	thresholds $Q$ and $M$. 

	Evolution is modelled through a stochastic birth-death process
		\cite{nowak2004emergence}, considering a well-mixed finite population of $Z$ 
			individuals whose members randomly form groups of size $N$ that play the game
			previously described. The fitness of each individual is computed averaging over all the possible group configurations she can be part of. Then, randomly selected individuals may adopt the strategy of  other random members of the
		population with a probability that increases with their fitness
		difference, amplified or reduced by the intensity of selection $\beta$~\cite{traulsen:2006a} (see \EM in \SM for details).
	Given the large number of strategies, we adopt the
		so-called rare mutation limit, which allows one to conveniently describe the
		prevalence of each strategy as a reduced Markov chain  \cite{fudenberg:2006, imhof2005evolutionary,
			hilbe2017memory, vasconcelos2017prl,omidshafiei2020navigating}. This approach enables the computation of the invasion diagram among
		all pairs of strategies, together with the prevalence of each
		strategy in the long-run. This
		framework also allows to compute an average cooperation
		level for different parameters of the model.

	\section{Results}
	\label{sec:results}
	
	\subsection*{Signalling versus reciprocity}
	
\begin{figure}[!t]
            			\centerline{\includegraphics[width=12cm]{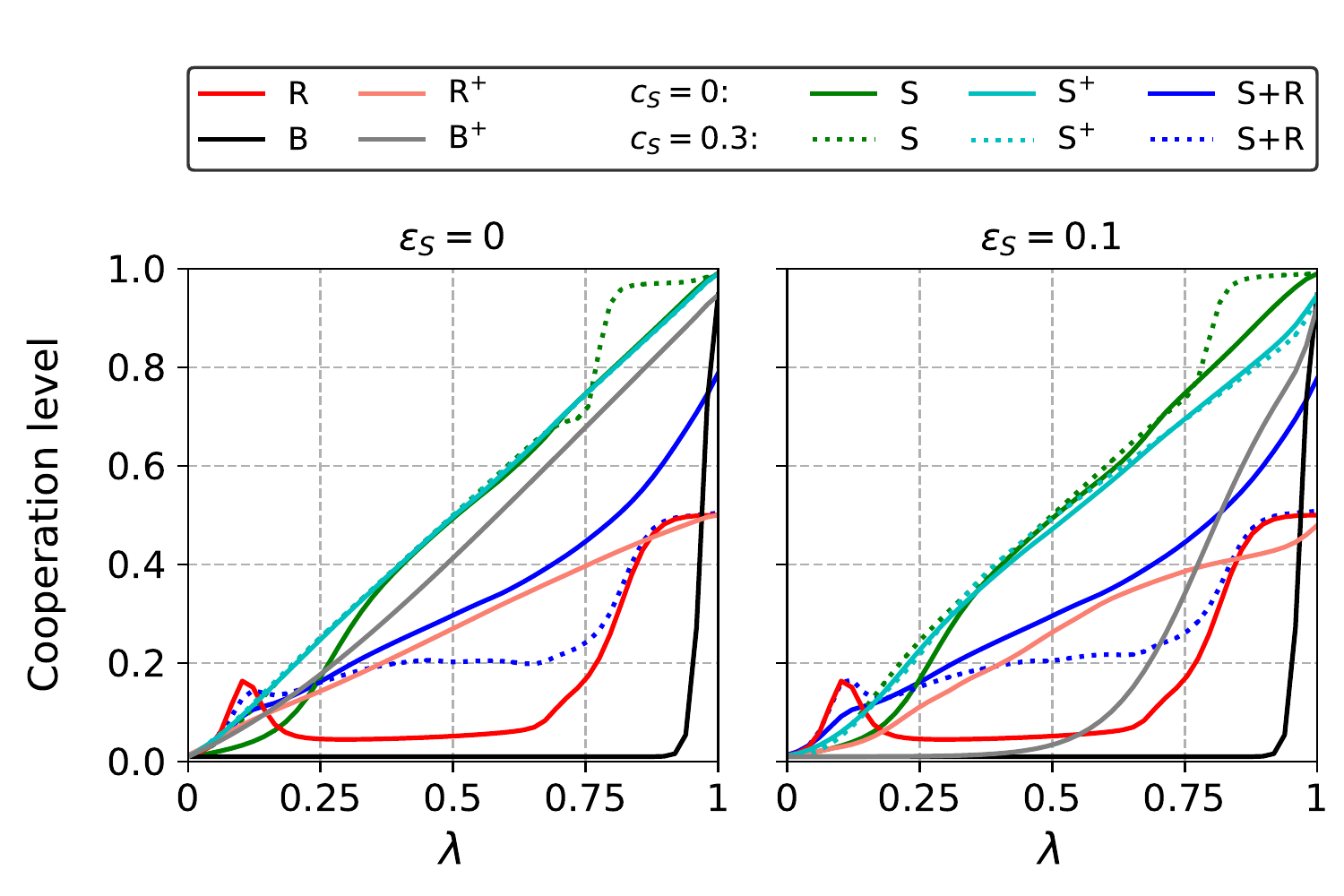}}
            	\caption{\textbf{Cooperation level attained under
            			different mechanisms.} We tested the following
            		mechanisms: reciprocity (\sR, 2 bits:
            		$[a_{M},a_{\tilde{M}}]$), signalling (\sS, 4 bits:
            		$[s_{G},s_{\tilde{G}};a_{Q},a_{\tilde{Q}}]$) and the
            		combination of signalling and reciprocity (\sSR, 6
            		bits:
            		$[s_{G},s_{\tilde{G}};a_{\tilde{Q},M},a_{\tilde{Q},\tilde{M}},a_{Q,M},a_{Q,\tilde{M}}]$). The
            		baseline scenario with only pure cooperators and defectors
            		is also included (\sB, 1 bit: $[a]$). For all mechanisms, strategies with the ability to
            		discriminate between $G$ and $\tilde{G}$ are also included: \sRp (4 bits:
            		$[a_{\tilde{G},\tilde{M}},a_{G,\tilde{M}},a_{\tilde{G},M},a_{G,M}]$), 
            		\sSp (6 bits: $[s_{G},s_{\tilde{G}};a_{\tilde{Q},G},a_{\tilde{Q},\tilde{G}},a_{Q,G},a_{Q,\tilde{G}}]$),
            		\sBp (2 bits: $[a_{G},a_{\tilde{G}}]$). Different values
            		of the cost of signalling are considered.
            		Cooperation levels have been computed for the combination
            		of both ecological conditions ($G + \tilde{G}$, as
            		determined by $\lambda$). We assumed $M=5$, $Q=N/2$,
            		$\omega=1$, $\beta=1$, $r=10$, $c=1$, $\epsilon=0.01$,
            		$N=9$, and $Z=100$.}
            	\label{fig1}
            \end{figure}

	We first study
          the cooperation level achieved through signalling (\sS) and
          reciprocity (\sR) both as separate and coexisting mechanisms
          (\sSR, see Table~S1 for a summary of the different combinations of
            mechanisms). Fig~\ref{fig1} shows
          the expected fraction of cooperative actions for those
          mechanisms together with a baseline (\sB) where only pure
          cooperation or defection is allowed. We also
            consider the case in which players can detect the current
            ecological condition---although with a probability of
            error $\epsilon_S$---and act accordingly (\sBp), possibly
            together with reciprocity or signalling (\sRp and \sSp,
            respectively, see also Table~S1). Our analysis is tailored to understand what
            mechanisms and combinations thereof result in the highest
            levels of cooperation as a function of the probability of
            encountering the profitable or unprofitable ecological
            conditions, determined by $\lambda$.

	In the absence of any mechanism (baseline condition
	\sB), defection prevails for almost the entire spectrum
	of $\lambda$.  This is because the public good game in $G$ is only
		marginally advantageous with the selected parameterisation, and
		therefore cooperation is observed only for $\lambda\approx 1$. By
	providing individuals with the capacity to reciprocate prior outcomes
	(condition \sR), the overall cooperation level improves,
	especially
		when favourable ecological conditions are frequent.  A much more
	effective mechanism is signalling alone (condition \sS),
	not only when it comes without any cost, but also for a moderate one
	($c_s = 0.3$). Finally, allowing individuals to use both signalling
	and reciprocity (condition \sSR) does not add value to the former: it
	diminishes the virtue of signalling, reducing the cooperation level to
	values between those obtained from signalling and reciprocity considered as separate mechanisms.This result
	is valid even in the presence of costly signals, in which case the
	\sSR strategies attain a cooperation level just above the one
	scored by \sR.
	
	Cooperation increases when individuals discriminate accurately between
	the state $G$ and $\tilde{G}$, \ie when errors in perception of the
	ecological condition are neglected
	($\epsilon_S=0$).  In the case of the baseline
	condition, the difference between \sB and \sBp is
	significant, the latter reaching cooperation levels almost as
	high as the signalling mechanism. When reciprocity is
		considered, the cooperation levels for \sRp also increases over \sR,
		but remains lower than 0.5, attained when $\lambda=1$.
	Finally, when signalling is considered, higher cooperation is
		registered when individuals can decide how to act as a function of
		the ecological condition (\sSp), increasing over the already good
		cooperation achieved by simple signalling (\sS) even for small
		$\lambda$.
	The main reason behind the higher cooperation observed when
		individuals are able to discriminate the ecological condition
		resides in the ability to conditionally defect when in $\tilde{G}$
		while developing a more cooperative strategy when in $G$, as long as
		this condition is not too rare (see some examples illustrated in
		Table~S2).
              Nonetheless, when individuals fail to properly
              discriminate between ecological conditions
              ($\epsilon_S>0$), the conditional strategies emerging in
              \sBp, \sRp, and \sSp are not that successful any
              more (see the right panel in Fig~\ref{fig1} for
              $\epsilon_S=0.1$ and with more detail in Fig~S1).
              The cooperation levels for the baseline condition
                and for the one entailing reciprocity are strongly affected by
              perception errors, whereas signalling (especially
              when costly) is able to preserve a high cooperation
              level. In other words, signalling represents a
                mechanism capable of correcting the individual
                perception errors, owing to the aggregation of
                information from multiple individuals.

	One may argue that the cooperation level is not always a fair measure
	of collective action, given that there is no need to cooperate in the
	case of abundance (i.e., when in $\tilde{G}$) as this does not lead to
	additional benefit. Indeed, when the discrimination between states
		is possible, conditions \sBp, \sRp, and \sSp all lead to defection in
		$\tilde{G}$, as discussed above. Fig~\ref{fig1} also shows that the
	cooperation level attained when only signalling is allowed is roughly proportional to the frequency of $G$ as
	given by $\lambda$, indicating that signalling favours cooperation
	mainly when the ecological condition entails a public good game. This
	is due to the possibility to distinguish the state $G$ from
	$\tilde{G}$ and to conditionally cooperate as a function of the
	signalling quorum---if reached---in spite of the coordination problems
	that can arise when interpreting the meaning of signals. This
	hypothesis gets support from Fig~S1, where we compute the cooperation
	level only for the state $G$: signalling strategies mostly cooperate
	in $G$ and do not cooperate when in $\tilde{G}$. Strategies that
	only use reciprocity or baseline mechanisms, instead,
	cannot act conditionally on the state $G$.
	
	Owing to these results, in the following we focus on the simpler
		conditions \sB, \sR, \sS, and \sSR, as the additional ability to act
		as a function of the ecological condition does not provide
		substantial advantages in the presence of perception error. We focus
		on the emergence and evolution of the behavioural strategies to
		understand why reciprocity jeopardises the benefits provided by
		signalling when the two mechanisms are jointly enabled.

	\subsection*{A new ecology of opportunistic strategies}
	
	The low cooperation observed in the presence of both signalling
		and reciprocity demands for an explanation. Why the co-existence of
	mechanisms that should promote cooperation not only appears
	unproductive but also jeopardises the ability to cooperate?  Why is
	reciprocity not just selected out in favour of strategies that use
	signalling, if that is advantageous?

\begin{figure} [t!]
		\centerline{\includegraphics[width=12cm]{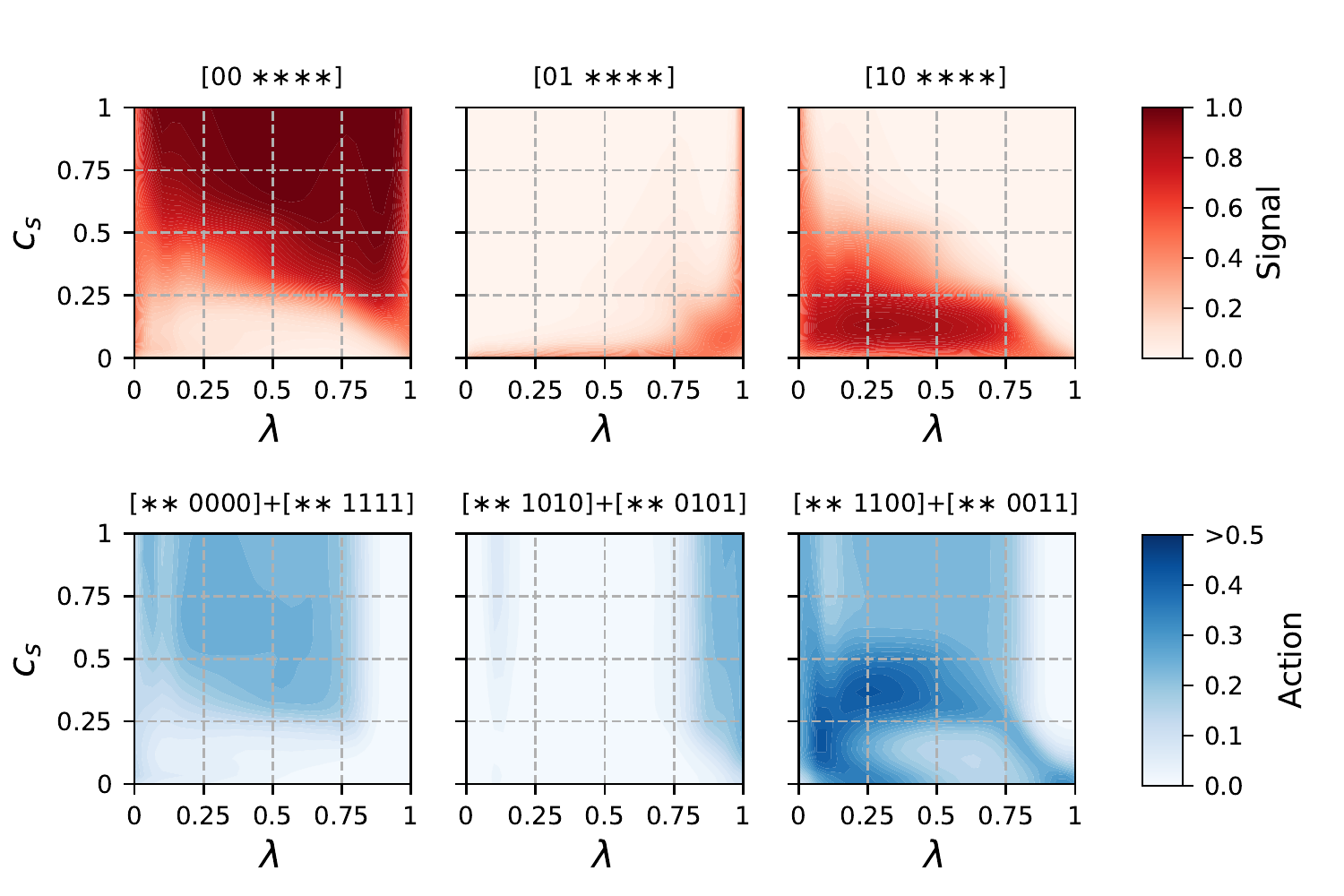}}
		\caption{\textbf{Prevalence of strategies grouped by signalling or acting
				behaviour}. The plots show the aggregated probabilities in the
			stationary distribution of strategies across the $\{\lambda,c_S\}$
			parameter space. In the top row, the signalling behaviour is
			considered, while the acting behaviour is displayed in the bottom
			row. Signalling strategies are grouped by the first two bits
			($s_{G},s_{\tilde{G}}$), ignoring the always-signalling group
			which has negligible prevalence. Concerning the action part, in
			the bottom row we show groups of strategies that exploit either
			reciprocity (middle panel) or signalling (right panel), in
			comparison to strategies that do not use any mechanism (left
			panel).  We assumed $M=5$, $Q=N/2$, $\omega=1$, $\beta=1$, $r=10$,
			$c=1$, $\epsilon=0.01$, $N=9$, and $Z=100$.}
		\label{fig2}
	\end{figure}

	In Fig~\ref{fig2}, we show for the \sSR conditions the prevalence of
	strategies that resort to signalling (first row) and direct
	reciprocity (second row) for different values of $c_{S}$ and
	$\lambda$. If we look at the prevalence of strategies grouped by their
	signalling behaviour 
	(top row), we note that signalling is significantly used whenever the
	cost is low.  Costly signals are used in situations where cooperation
	is required, i.e., in the $G$ state ($s_{G}s_{\tilde{G}}=10$, see the
	top-right panel), but also to identify a rare condition
		$\tilde{G}$ ($s_{G}s_{\tilde{G}}=01$ and high $\lambda$, see the
		middle-top panel).
	Moreover, the emission of signals does not lead to strategies that
	just exploit such information, ignoring previous interactions
	(see Fig~\ref{fig2}, bottom-right panel).
	Response to signals that ignore memory do not dominate, and neither
	pure reciprocal strategies do (see Fig~\ref{fig2}, bottom row). Overall, pure signalling strategies are
	not improved but jeopardised by reciprocity. A similar situation occurs
	for other values of $M$ (see Fig~S2).

	To fully understand the nuances of the interaction between signalling
	and reciprocity, we analyse the dynamics among the individual
	strategies and their emergence and dominance in an evolutionary
	context. We found that, despite
	the large strategy space, one can conveniently cluster strategies
	in a few drifting groups, i.e., strategies that are neutral
	among each other (see \EM in \SM for details). These drifting groups are identified by distinct colours in
	Fig~\ref{fig3}, being characterised by the following properties:
	
		\begin{figure}[!t]
		\centering
		\includegraphics[width=\textwidth]{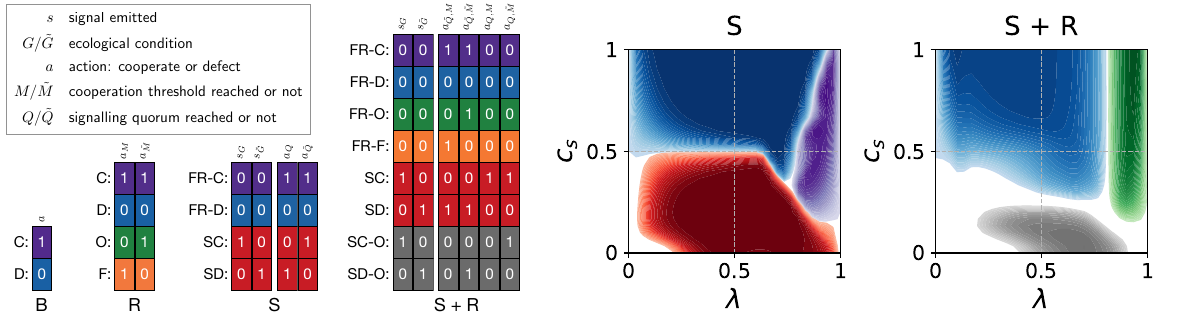}
		\caption{\textbf{Main groups of strategies.} \emph{Left}: Relevant
			strategies are displayed for each possible setting, including the
			baseline setting where no mechanism is involved. Each labelled
			group is represented by a prototype strategy, and includes mutant
			strategies that drift with the prototype (see
			\EM in \SM for
			details). Strategies are colour-coded in a consistent way
			indicating coherent resulting behaviours (e.g., red represents
			those strategy groups that exploit signalling only). \emph{Centre
				and right}:
			Stationary distribution of the drifting groups of strategies in the
			$\{\lambda,c_S\}$ parameter space when only signalling (\sS, centre) and
			reciprocity combined with signalling (\sSR, right) are included. The
			same colour-coding as in the left panel is used. Colour darkness
			indicates the intensity of the probability in the stationary
			distribution. Only probabilities higher than $0.5$ have been
			displayed to facilitate the visualization in overlapping areas (see also Fig~\ref{fig4}).
			Parameters assumed: $M=5$, $Q=N/2$, $\omega=1$, $\beta=1$,
			$r=10$, $c=1$, $\epsilon=0.01$, $N=9$, and $Z=100$.  }
		\label{fig3}
	\end{figure}

	\begin{description}
		\item[Unconditional strategies (\sB)] When no mechanism is used, like
		in the baseline condition, the relevant strategies are just
		\emph{unconditional cooperation} (\sC) and \emph{unconditional
			defection} (\sD).
		\item[Reciprocity-based strategies (\sR)] When reciprocity is present,
		four different strategies are possible, as determined by two bits
		($a_{M},a_{\tilde{M}}$). Besides pure cooperation (\sC) and pure
			defection (\sD), we count a strategy that only
		cooperates if the threshold $M$ was reached and defect
		otherwise---referred to as \textit{follower} (\sF)---and the
		opposite strategy, which cooperates only when the threshold was not
		reached---referred to as \textit{opportunistic} (\sO). The former
		can also be seen as an N-person analogue of Tit-for-Tat, and the
		latter as the analogue to Anti-Tit-for-Tat, a compensating strategy that only cooperates when the others
		refuse to do so~\cite{baek2016comparing, do2017combination,domingos2020timing}.
		\item[Signalling-based strategies (\sS)] When only signalling is used,
		relevant strategies must consider both the signalling and the action
		components (hence four bits:
		$s_{G},s_{\tilde{G}},a_{\tilde{Q}},a_{Q}$ and $2^4$ possible
		strategies). In this case, we found four well-definite drifting
		groups, illustrated in Fig~\ref{fig3} with a
		prototypical example for each case. Besides those strategies that
		free-ride the cost of signalling $c_s$ and just cooperate and
		defect---referred to as \sFRC and \sFRD, respectively---we observe
		two strategies that signal to indicate when it is better to
		cooperate or defect as a function of the ecological conditions: \sSC
		signals to cooperate when in $G$, while \sSD signals to defect when
		in $\tilde{G}$. Both obtain the same net effect, although \sSC
		prevails over \sSD for most values of $\lambda$ because paying the
		cost $c_S$ is better tolerated when a benefit is expected.
		\item[Strategies based on Reciprocity and Signalling (\sSR)]
		Finally, when both reciprocity and signalling are present, the full
		set of $2^6$ strategies is available. Nonetheless, as before, only a
		limited number of drifting groups dominate (see list in
		Fig~\ref{fig3} and \SM). As
		before, the possibility to reciprocate leads to the
		\emph{opportunistic} and \emph{follower} strategies (\sFRO and
		\sFRF, respectively), and the possibility to use signals leads to
		the observed strategies that signal to cooperate or defect (\sSC and
		\sSD). More importantly, two additional groups emerge that were not
		previously found, and that are responsible for the dismiss of
		cooperation when signalling and reciprocity co-exist. These are {\it
			opportunistic} strategies that signal either in state $G$ or
		$\tilde{G}$, and cooperate accordingly to the signal they emit, but
		only when the cooperation threshold $M$ was not reached in the
		previous round. We refer to these strategies as \sSCO and \sSDO.
	\end{description}

	The identified groups allow to appreciate the prevalence of different
	types of strategies when the available mechanisms change. Most
	importantly, we note that opportunistic strategies dominate in a large
	portion of the parameter space when both signalling and reciprocity
	are present (condition \sSR), as shown in the right
	panel of Fig~\ref{fig3}. Specifically, \sSCO and \sSDO dominate
	in those regimes where, in the absence of reciprocity, we would have
	observed \sSC and \sSD strategies (see condition \sS, center panel
		in Fig~\ref{fig3}). Indeed, opportunistic strategies
	resort to both information from signals and from the previous
	round to opportunistically cooperate either in $G$
	or in $\tilde{G}$, leading to lower overall levels of
	cooperation. Similarly, the reciprocal strategy \sFRO in \sSR
	replaces \sFRC in \sS, when only signalling is available.  Since
	\sFRO pays fewer costs by cooperating opportunistically, it also
	reduces the overall levels of cooperation.

\begin{figure}[!t]
	\centerline{\includegraphics[width=12cm]{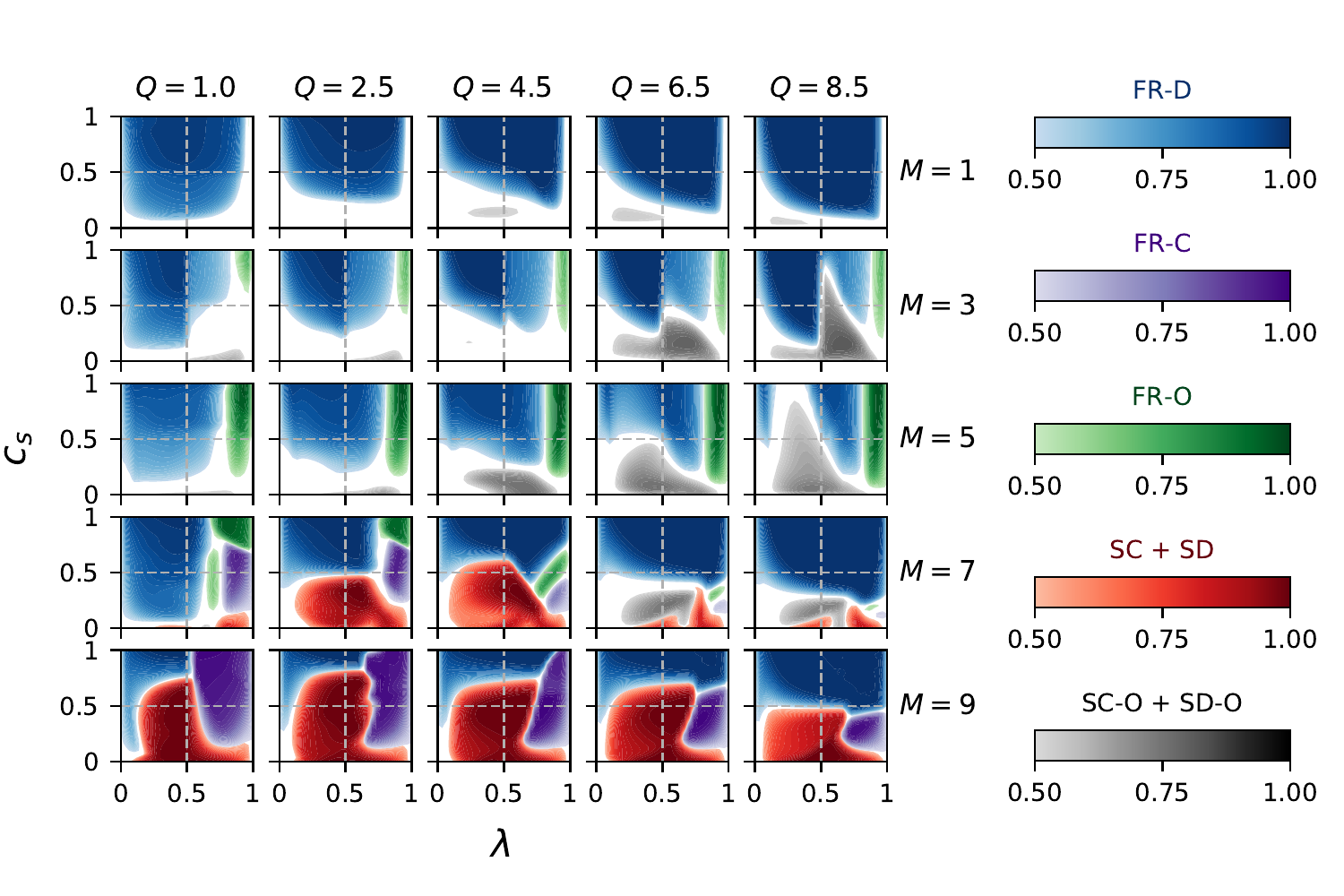}}
	\caption{\textbf{Stationary distribution in the \sSR
			condition}. Drifting groups of strategies are displayed across
		the $\{\lambda,c_S\}$ parameter space for all values of thresholds
		$Q$ and $M$. Only probabilities larger than
		$0.5$ have been displayed to facilitate the visualization in
		overlapping areas.  Other parameters: $\omega=1$, $\beta=1$,
		$r=10$, $c=1$, $\epsilon=0.01$, $N=9$, and $Z=100$.}
	\label{fig4}
	\end{figure}

	Naturally, the prevalence of opportunistic strategies 
	depends on the game parameters. The complete overview of the
	dominating strategies is provided in Fig~\ref{fig4} (see also Fig~S3
	for conditions \textbf{B}, \textbf{R} and \textbf{S}), which reveals
	how different groups may dominate in different portions of the
	parameter space when varying the signalling quorum $Q$ and the
	cooperation threshold $M$. Generally speaking, we observe that
	opportunistic strategies are more important when the requirements for
	cooperation are mild ($M\leq5$), while higher
	requirements 
	entail larger usage of signals, even when the related cost is non
	negligible. The opportunistic strategies undermine the achievement of
	high cooperation levels, and can be counteracted only by rising the
	requirements for cooperation to $M\geq7$. Under such conditions, a
	large portion of the parameter space is dominated by pure signalling
	strategies or unconditional cooperators \sFRC that free ride on the
	signals emitted by others, the latter appearing only when the
	ecological conditions entail a frequent public good game (high
	$\lambda$).  Note that the drifting groups of strategies we have
		identified are evolutionary robust for a wide range of parameters
	(see also \SM). They are generally not invaded
	by any other strategy with a probability higher than the one obtained
	through neutral drift, and when that happens it occurs with a
	probability much lower than the invasion in the opposite direction
	(see invasion graphs in Fig~S4 and preferred
          directions of invasions~\cite{fudenberg:2006, imhof2005evolutionary,kurokawa2009emergence,
			vasconcelos2017prl,omidshafiei2020navigating} in the
          Section S4 of the \SM). These results
	are robust to variations in the number of rounds---as shown in Fig~S5,
	where the probability to play an additional round of the iterated game
	is set to $\omega=0.9$---and to the existence of different type of
	errors, such as the inability of individuals to correctly discriminate
	the ecological condition ($\epsilon_S>0$, see Fig~S6).

	\subsection*{Alternative scenarios}
	
\begin{figure}[!t]
	\centerline{\includegraphics[width=12cm]{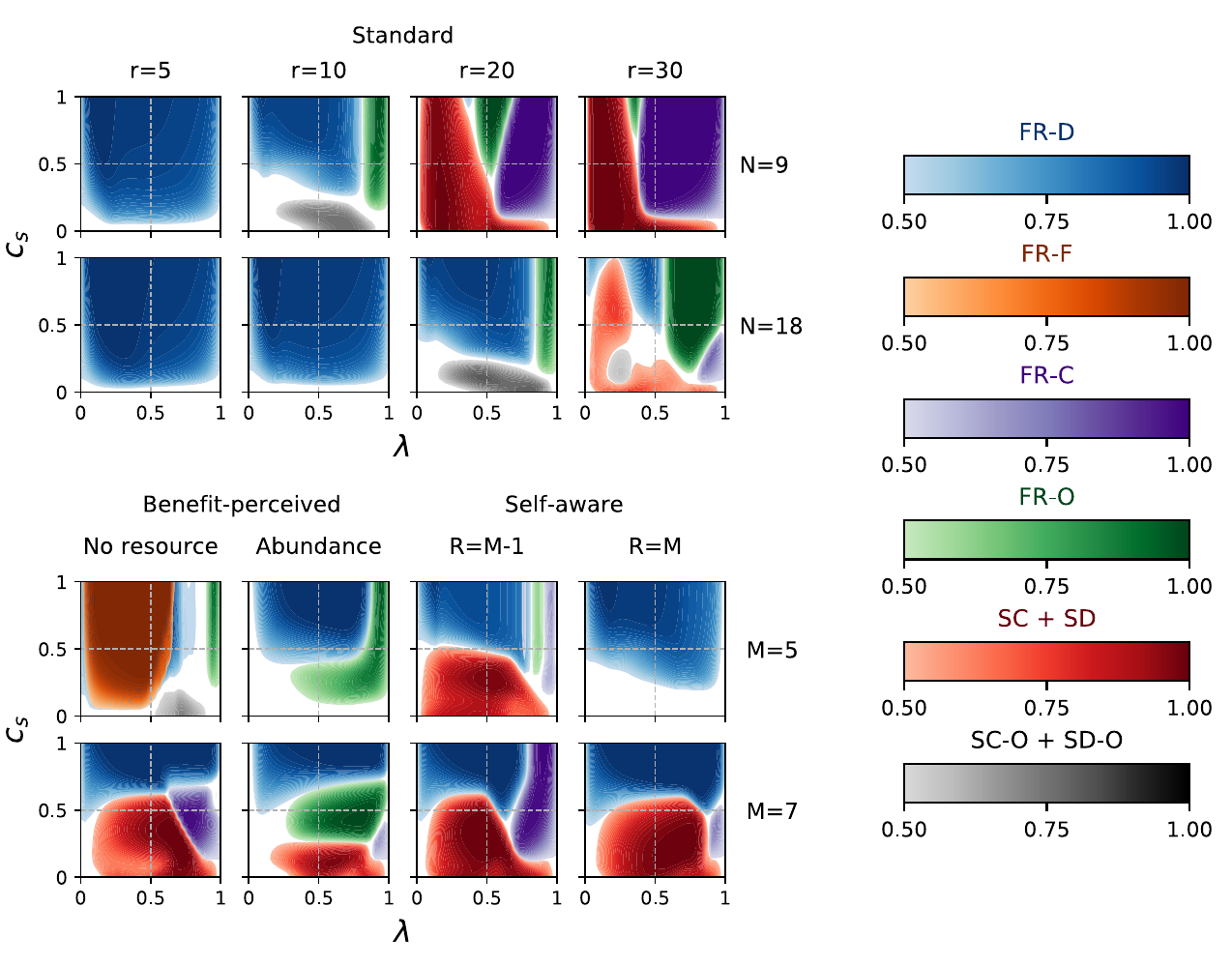}}
	\caption{\textbf{Stationary distribution of the main groups of
			strategies for different scenarios.} \emph{Top panels}: Effects
		of different group sizes $N$ and multiplication factor $r$ in the
		standard scenario, with $M=\frac{5}{9}N$. Note how, within the
		panel corresponding to $N=18$ and $r=30$, one can see a clear
		transition between \sSCO to \sSC. \emph{Bottom panels}: Effects of
		the benefit-perceived and self-aware scenarios, here with standard
		parameterisation (see also \SM for additional cases). In the
		abundance case, \sSC corresponds to [10\,0$\ast$11] while \sSD to
		[01\,110$\ast$]. In the self-aware scenario, when $R<5$, \sSC
		corresponds to [10\,001$\ast$], \sSD to [01\,1$\ast$00] and \sFRC
		to [00\,1$\ast$$\ast$$\ast$].
		For $R\ge 5$, groups of strategies remain as in the standard
		scenario. Other parameters assumed unless otherwise indicated:
		$Q=N/2$, $\omega=1$, $\beta=1$, $r=10$, $c=1$, $\epsilon=0.01$,
		$N=9$, and $Z=100$.} 
	\label{fig5}
\end{figure}

	The results discussed so far are valid also for other multiplication factors $r$ and group sizes $N$. In Fig~\ref{fig5}, we show the
	influence of both for intermediate values of $M$ and $Q$. 
	As expected, low multiplication factors lead to the dominance of defection
	in the whole parameter space, while higher values promote cooperative
	strategies, and signalling is exploited widely when $\lambda<0.5$,
	i.e., when the $G$ state is rare. In that case, signals are used to
	identify the need for cooperation when in $G$, hence
	avoiding to pay costs with unproductive cooperation when in the
	$\tilde{G}$ state.

	Increasing the size of the groups ($N$) has a similar effect to reducing
	the multiplication factor, \ie prevalence of less cooperative strategies \cite{boyd:1988}, as can be noted by the similar pattern
	between the case with $r=10$ and $N=9$ and the case with $r=20$ and
	$N=18$ in Fig~\ref{fig5}. This effect has been also
	identified in other types of non-linear
        N-person games~\cite{pacheco2009evolutionary,santos2011risk,souza2009evolution},
  	and when the only available mechanism of cooperation was
          either signalling~\cite{pacheco2015co} or reciprocity \cite{van2012emergence,pinheiro2014evolution}. This dependence, however, may be influenced by how strategies are encoded \cite{takezawa2010revisiting}.

Overall, the results of Fig~\ref{fig5} confirm that signalling strategies emerge
	when cooperation is most needed \cite{pacheco2015co}, and that mild requirements for
	cooperation lead to the emergence of signalling strategies that
	however cooperate opportunistically. Indeed, our results suggest that information about the outcome of previous games is 
	promoting opportunistic behaviours. This can however be an effect of
	the absence of a precise feedback on the outcome of the game, as the
	cooperation threshold $M$ can be achieved even in the $\tilde{G}$
	state, when there is no benefit to compensate the costs of
	cooperation, hence justifying an opportunistic behaviour that
	cooperates only half of the times. We therefore tested an alternative
	scenario in which individuals can base their decision on the actual
	reception of a benefit in the previous round. We refer to this
	scenario as the benefit-perceived scenario. This affects only the
	behaviour in the $\tilde{G}$ state, whether the benefit is
	unconditionally provided (\emph{abundance}) or unconditionally not
	{provided} (\emph{no resource}, see \EM in \SM) for
	details). Fig~\ref{fig5} and Fig~S7 show that when $\tilde{G}$ is frequent ($\lambda<0.5$)
	but returns no benefit, and when requirements for cooperation are
	mild ($M=5$), the follower strategies \sFRF dominate. These
	strategies use memory to identify the $G$ state---the only possibility
	to obtain a benefit---without the need of signals. Actually, \sFRF
	uses knowledge about the obtained benefit as a cue that identifies the
	ecological condition $G$, and cooperates accordingly. This mechanism
	only works if the group is able to reach the threshold $M$ at some
	point, reason why this group of strategies disappears for high values
	of $M$.
	Conversely, in case of abundance, pure opportunistic strategies \sFRO
	take over a large portion of the parameter space, also dominating
	signalling strategies (see Fig~\ref{fig5} and Fig~S8). Opportunistic strategies can gain in both
	the $G$ and $\tilde{G}$ state without paying a cost for signalling,
	but obtain a benefit only half the times when in $G$. For this reason,
	\sSC and \sSD still dominate when the costs of signalling $c_s$ is low
	and $\lambda$ sufficiently high, as signalling allows to
	systematically cooperate in $G$ without incurring in errors.

	Another reason for the emergence of opportunistic strategies may come
	from the fact that agents are not aware of the importance of their
	contribution towards reaching the cooperation threshold $M$, as the
	information about how many other agents cooperated in the previous
	round is not explicitly available. Therefore, attaining $M$ (when this
	is mild) may well have been possible without the own contribution,
	hence promoting opportunistic choices. We therefore tested what
	strategies would emerge in such a \emph{self-aware scenario} in which
	reciprocity is based on the threshold $R$ of individuals other than
	the focal player that cooperated in the previous round (see
        \EM in \SM
	for details). Fig~\ref{fig5} and Fig~S9 show, interestingly, that opportunistic strategies are
	nearly completely wiped out in favour of unconditional defection when
	$R \geq M$, or of strategies using signals when $R < M$.  Indeed, as
	$R$ is now partially unrelated from the attainment of benefit in $G$,
	signalling strategies are more reliable and dominate largely over
	opportunistic or mixed strategies. Overall, we see that reducing the
	requirements for group reciprocity actually favours the evolution of
	signalling strategies.

\section{Discussion and Conclusions}
\label{sec:discussions}

In a complex collective action problem in which different and
contrasting ecological conditions may be encountered, the interplay
between mechanisms known to individually promote cooperation, such as
signalling and reciprocity, is far from trivial.  Our results suggest
that successful strategies in the realm of direct reciprocity are not
selected in the presence of signalling. Even free-riders, and
particularly those that do not pay the cost of emitting signals, make
use of the information supplied by others to decide how to act. As a
result, reciprocity is relegated to a narrow range of the possible
environmental conditions, and when it appears, it does it in
combination with signalling. The latter mechanism, on the other hand,
proves more reliable owing to the ability to support the collective
discrimination of the ecological context, which is key to support
cooperation especially when the problem requirements are tight,
being even able to correct errors in the perception on the
	ecological conditions.

Interestingly, these results are compatible with observations that
show how signalling is extensively preferred over reciprocity in
non-human living beings~\cite{hammerstein:2003b}. Indeed, it was
argued that reciprocity is rare because of cognitive limitations
\cite{stevens2004nice,stevens2005evolving,hauser2009evolving,moreira2013individual},
ecological constraints~\cite{whitlock2007costs}, and an evolutionary
bootstrapping problem~\cite{andre2014mechanistic}.  Particularly,
reciprocity requires an important degree of complexity at the individual level (cognitive skills that allow
memory), whereas signalling benefits from the fact that it does not
need a large set of repeated interactions to emerge, only relying on
the quorum of signals within each group. As already suggested, quorum
sensing can be used to precisely identify profitable ecological
contexts, whereas reciprocal strategies base their success only on the
feedback from others, being successful only on average and in the long
run. Additionally, reciprocity hardly evolves when the available
  information is  incomplete~\cite{kurokawa2016evolutionary}.

Another reason for the lower prevalence of reciprocity has
been identified in the complexity of the ecology of strategies
emerging from its combination with signalling.
When both signalling and reciprocity are possible, we did not find any
synergy, and lower levels of cooperation are reached when compared to,
e.g., situations in which only signalling is available. This result
remains valid even in the presence of costly signals.
In fact, owing to the additional degrees of freedom granted by signals
and associated responses in combination with memory-based actions, a
new class of opportunistic strategies emerges that prevails in a wide
range of parameters, wherein individuals free-ride on the efforts (and
signals) of others, but only when their contribution is not
decisive. This result highlights the subtle interplay between
individuals' strategic complexity and cooperation, suggesting that
evolution tends to select simpler heuristics to foster
pro-sociality. This surprising result fits nicely with recent studies
in the realm of reputation-based systems 
\cite{santos2018social}, where simple moral principles are shown to maximize cooperation, even in complex environments.

Given the size of the strategy space emerging from the combination of
signalling and reciprocity, further intricacies where intentionally
left out from our model. We assume, for instance, that signalling can
come with a cost, whereas memory is cost-free. Here, memory is seen as
an intrinsic feature that is not activated deliberately by the
individual. However, our results show that reciprocity emerges in a
minimal number of scenarios even with cost-free memory; thus, an
additional cost would further reduce the chances of reciprocity-based
strategies. One could also consider more complex signals~\cite{de2020interplay}, individual intentions~\cite{han2015synergy}, and reciprocal strategies that react differently for each configuration
  of the group in terms of the cooperation level observed in the previous
  round, for instance, discriminating between a few and many
  cooperators~\cite{pinheiro2014evolution,kurokawa2018evolution}. Such additional feature
  could reveal more complex dynamics for signalling and reciprocity to
  possibly act in synergy. Future work can explore these interesting
  paths.

Overall, our framework provides novel insights into the analysis of
behavioural dynamics in the presence of multiple cooperation
mechanisms, showing how opportunistic behaviour can arise within a
complex ecology of behavioural strategies. Under demanding pressures
from the collective action problem, our results put forward further
explanations on the prevalence of signalling and quorum dynamics as a
ubiquitous property in nature, showing how it may prevail even when in
direct competition with other cooperation mechanisms that benefit from
higher individual cognitive skills.

\section*{Author Contributions}
LAMV implemented the model. All authors designed the research, analysed the results, wrote and revised the manuscript, gave final approval for publication, and agreed to be held accountable for the work performed therein.

\section*{Funding}
This work was partially supported by the project DICE (FP7 Marie Curie Career Integration Grant, GA: 631297).
L.A.M.-V. also acknowledges the support of the European Research Consortium for Informatics and Mathematics through an Alain Bensoussan Fellowship. F.C.S. acknowledges the support by FCT-Portugal through grants UIDB/50021/2020, PTDC/EEI-SII/5081/2014, and PTDC/MAT/STA/3358/2014. V.T. and F.C.S. acknowledge partial support from the project TAILOR (H2020-ICT-48 GA: 952215).

\end{document}


\maketitle

In this Supplementary Material we further elaborate on the methods and results
discussed in the main text. In Section S1, we provide an additional
  details on the analytical and numerical procedures employed. In Section S2, we detail the nature of the strategies in
each drifting group. We show further
evidence of the robustness of our conclusions in Section S3, providing extended analyses and discussion of the whole parameter space. Finally, in Section S4 we deduce analytically the conditions under which some significant strategies are more likely to be invaded by others.

\section*{S1. Extended Methods}

As detailed in the main text, we consider a finite population of $Z$ individuals, which form
	random groups of size $N$. Within each group, individuals interact
	through a non-linear and stochastic $N$-player iterated game. 
	As we detail in the following
	sections, we present a finite population model of evolution, in which
	the proportion of time spent at equilibria can be explicitly
	calculated. We consider that individuals revise their behaviour by
	social learning, such that individuals with higher fitness will tend
	to be imitated more often. Here, fitness is given by the average
	payoff obtained from a large number of $N$-player interactions. Given
	the large number of different strategies ($2^6$ for \sSR), our analysis can be simplified by
	adopting the limit of rare mutations, corresponding to maximal
	reduction of configurations of interest in a hierarchy of possible
	approximations~\cite{fudenberg:2006, imhof2005evolutionary, hilbe2017memory,
		vasconcelos2017prl}. Under a process of imitation
	dynamics, only a new mutation can introduce a new strategy and,
	whenever mutations are rare, new strategies will either invade a
	resident population or become extinct. Thus, assuming that mutations
	take place much slower than the reproductive or imitation dynamics, as
	we describe later, there exist a maximum number of two strategies $A$
	and $B$ at the same time in the population. We denote as $N_\Lambda$ and
	$Z_\Lambda$ the number of individuals of type $\Lambda\in\{A,B\}$ in the group and
	population, respectively, such as $N_A+N_B=N$ and $Z_A+Z_B=Z$.

	\subsection*{Game payoffs and strategy space}
	
	We first assume that the ecological
	context is in one of two states $L$: a public
	good game state $G$ with probability $\lambda$ and a non public good
	game $\tilde{G}$ with probability $1-\lambda$. In each round,
	individuals may decide to contribute a value to a common pool (to
	Cooperate, C) or refuse to contribute (to defect, D). In the $G$
	state, a collective benefit is produced to the extent that at least
	$N_C\geqslant M$ individuals contributed to the public good, where $M$
	represents a coordination threshold. In this case, each member of the
	group obtains a benefit $b=rc\,N_C/N$, with $c$ the cost of
	cooperating, $N_C$ the number of cooperators, and $r>1$ the
	multiplication factor. If $N_C<M$ no collective benefit is produced,
	and all receive 0.
	
	In \sSR, behavioural strategies represent under what condition
	individuals will act in one or another way. We assume that individuals
	make two decisions within each round of the game:
	\begin{itemize}
		\item {\it Signalling}. First individuals decide if they signal
		($S$) or not ($\tilde{S}$). Signalling comes with a cost $c_S$. This
		decision is only based on the current ecological context, hence it
		is encoded by two bits in the strategy space $s_\alpha$, with
		$\alpha\in\{G,\tilde{G}\}$. If more than a given number $Q$ of
		individuals in the group signal, we say that the quorum was reached
		and the group is signalling.
		\item {\it Acting}. After the signalling phase, individuals decide
		whether to cooperate ($C$) or defect ($D$). Cooperation comes with
		an individual cost $c$, whereas defecting is free. In this phase,
		the decision is made taking into account if the group reached the
		signalling quorum and if at least $M$ individuals cooperated in the
		previous round. Therefore four bits characterise this decision in
		the strategy space $a_{\gamma,\gamma'}$, with
		$\gamma\in\{Q,\tilde{Q}\}$ and $\gamma'\in\{M,\tilde{M}\}$ representing
		all combinations from the achievement---or not---of the signalling
		quorum ($\gamma$) and the attainment---or not---of the cooperation
		threshold ($\gamma'$). Note that in the count of individuals that
		cooperated is also included the focal player; we also analysed the
		scenario where the focal player excludes herself from this count
		(see below).
	\end{itemize}
	In summary, strategies are given by six bits
	$\{s_{G}, s_{\tilde{G}}; a_{\tilde{Q},M}, a_{\tilde{Q},\tilde{M}},
	a_{Q,M}, a_{Q,\tilde{M}} \}$.
	We restrain our analysis to pure strategies, \ie
	$s_\alpha,a_{\gamma,\gamma'}\in\{\epsilon,1-\epsilon\}$, where
	$\epsilon$ represents the small probability that an individual commits
	an error, that is, chooses the opposite option that she intends to. In
	order to make easier the notation we will write strategies just using
	0 and 1 instead of $\epsilon$. For example, an individual playing the
	strategy $[100011]$ signals only when in $G$ and cooperates only if
	the group was signalling (no matter if the threshold $M$ was reached
	in the previous round); on the contrary an individual with the
	strategy $[001010]$ never signals but cooperates if enough individuals
	$M$ cooperated in the previous round, neglecting signals.
	
	The payoff that each individual obtains in a given round of the
	repeated game is the sum of the benefit (if any) and the costs of
	signalling and cooperating (if she decides to do so).  Since a
	game is played for multiple rounds, the action of a player in a
	round may be influenced by the actions of the group in the previous
	round. If we consider two strategies, A and B, the iterative
	process can be described as a Markov chain with a
	stochastic matrix $\mathbf{A}$ whose elements ${A}_{ij}$ represent the
	transition probabilities between states $i$ and $j$ in a group formed
	by $N_A$ and $N_B$ individuals. Each state represents the number of
	individuals of each type that cooperates in a given round
	$\left(N_{C,A}(i),N_{C,B}(i)\right)$ (with $N_C(i)=N_{C,A}(i)+N_{C,B}(i)$) and then
	the parameter space can be written as
	$\{0,\dots,N_A\}\times\{0,\dots,N_B\}$.

	Given the stochastic nature of this repeated game and errors associated, for a large number of rounds (\ie the
		probability of playing a new round is $\omega=1$), initial
	conditions are not important since all possible states are
	eventually visited. In this situation, the probability $\mathbf{p}^L$ that the system is found in
	state $\mathbf{p}_{i}$ for a given ecological context
	$L$  just corresponds
	to the normalised eigenvector associated with the first eigenvalue of
	$\mathbf{A}^L$~\cite{sigmund:2010}.
	On the other hand, if the number of rounds are finite ($\omega<1$), the probability
      $\mathbf{p}^L$ is computed as follows:
	\begin{equation}
	\mathbf{p}^L= (1-\omega) \, \mathbf{p^0}\left( \mathbf{I} - \omega \mathbf{A}^L \right)^{-1}
	\end{equation}
	where $\mathbf{I}$ is the identity matrix and $\mathbf{p^0}$ the vector with the
	probabilities of the initial states. The length of each dimension of
	these matrices and vectors are $(N_A+1)(N_B+1)$.  We assume that the
	initial conditions in this game is
	${p^0}_{i}=\epsilon N_C(i)+(1-\epsilon)(N-N_C(i))$ independently of
	the ecological context. In other words, all individuals begin
	defecting (except for errors). This choice---although seemingly
	arbitrary---represents the case in which the group is presented with a
	new instance of the problem and has no prior knowledge
	to exploit. In such conditions, agents cannot rely on any memory about
	the outcome of previous interactions, equivalent to the state
	in which the cooperation threshold was not reached.
	
	The transition probabilities $A_{ij}$ can be expressed as
	\begin{equation}
	A_{ij}=Pr^A(i,j)Pr^B(i,j),
	\end{equation}
	where $Pr^\Lambda(i,j)$ stands for the probability that $N_{C,\Lambda}(j)$
	individuals cooperate out of the $N_\Lambda$ that are in the group,
	corresponding then to the probability mass function of the binomial
	distribution:
	\begin{equation}
	Pr^\Lambda(i,j)=\binom{N_\Lambda}{N_{C,\Lambda}(j)}P_{C}^\Lambda(i)^{N_{C,\Lambda}(j)}(1-P_{C}^\Lambda(i))^{N_\Lambda-N_{C,\Lambda}(j)},
	\end{equation}
	which is a function of the probability $P_{C}^\Lambda(i)$ that an
	individual following strategy $\Lambda$ cooperates when in state
	$i$. This probability depends on whether the thresholds $Q$ and $M$
	were reached and how the individual reacts according to her strategy:
	\begin{equation}
	\begin{split}
	P_{C}^\Lambda(i)&=\Theta\left(N_S-Q\right)\Theta\left(N_C(i)-M\right)\,a_{Q,M}^\Lambda \\
	&+\Theta\left(N_S-Q\right)\left[1-\Theta\left(N_C(i)-M\right)\right]\,a_{Q,\tilde{M}}^\Lambda \\  
	&+\left[1-\Theta\left(N_S-Q\right)\right]\Theta\left(N_C(i)-M\right)\,a_{\tilde{Q},M}^\Lambda \\  
	&+\left[1-\Theta\left(N_S-Q\right)\right]\left[1-\Theta\left(N_C(i)-M\right)\right]\,a_{\tilde{Q},\tilde{M}}^\Lambda, 
	\end{split}
	\end{equation}
	where $\Theta(x)$ stands for the Heaviside step function, \ie
	$\Theta(x)=1$ if $x\geqslant 0$ and $0$ otherwise. Note that
	$\mathbf{A}$ is different if individuals exclude themselves in the
	count of the threshold for reaching the cooperation threshold M (see
	below for an alternative).
	
	The average payoff per round that a $k$-strategist obtains in the
	group under the ecological context is 
	$W^L_\Lambda(N_A,N_B) = \mathbf{w}_{\Lambda}^L\cdot\mathbf{p}^L$, where
	$\mathbf{w}_\Lambda$ corresponds to the vector with the payoffs of each possible state for
	the player with strategy $\Lambda$:
	\begin{equation}
	{w}^L_{\Lambda,i}=rc\,\frac{N_C(i)}{N} \Theta(N_C(i)-M) - c\,\frac{N_{C,k}(i)}{N_\Lambda}  -c_S\,s^\Lambda_{L},
	\end{equation}
	
	Finally, the average payoff of an individual with strategy $\Lambda$ taking into account
	both ecological contexts is
	$W_\Lambda(N_A,N_B)=\lambda W^{G}_\Lambda(N_A,N_B)+(1-\lambda)
	W^{\tilde{G}}_\Lambda(N_A,N_B)$.

	\subsection*{Evolutionary dynamics}
	
	Once the average payoff per round $W_\Lambda(N_A,N_B)$ is obtained, we
	compute the average payoff $\Pi_\Lambda$ over all the possible group
	combinations \cite{hauert2006synergy,pacheco2009evolutionary,souza2009evolution,pinheiro2014evolution}:
	\begin{equation}
	\Pi_\Lambda=\sum_{N_\Lambda=1}^{\min(N,Z_\Lambda)} H(N_\Lambda-1,N-1,Z_\Lambda-1,Z-1) W_\Lambda(N_A,N_B),
	\end{equation}
	where the hypergeometric distribution can be expressed as
	\begin{equation}
	H(N_\Lambda,N,Z_\Lambda,Z)=\frac{\binom{Z_\Lambda}{N_\Lambda}\binom{Z-Z_\Lambda}{N-N_\Lambda}}{\binom{Z}{N}}.
	\end{equation}

	{
		In order to simplify the evolutionary dynamics analysis, we adopt the small mutation approximation~\cite{fudenberg:2006}. Under this approximation, whenever a mutant (an invasor) appears in a resident population, two possible final scenarios occur before any other mutation takes place: the mutant trait fixates in the population (it is imitated by all residents) or mutants are expelled (they imitate the resident strategy). In either case, the final state is monomorphic and no mixed state is considered, which could lead to a much more complex ecosystem of strategies~\cite{martinez:2012, vasconcelos2017prl}. The transition probability between pairs of strategies is determined as a fixation probability, \ie the probability that a single mutant with a strategy $j$ invades a population formed by $Z-1$ individuals that follow a strategy $i$~\cite{nowak2004emergence,fudenberg:2006, imhof2005evolutionary}: 
		\begin{equation}
		\rho_{ij}= \left( 1+\sum_{m=1}^{Z-1}\prod_{k=1}^m \frac{T^{-}(k)}{T^{+}(k)} \right) ^{-1},
		\label{eq:rho}
		\end{equation}
		where $T^{+}(k)$ is the probability that an individual of the resident
		strategy $i$ imitates a mutant  $j$ and $T^{-}(k)$ is the
		probability that an individual of the mutant strategy $j$ imitates a
		resident $i$ in a population of $k$ individuals playing the
		resident strategy.  Assuming a Fermi probability function for pairwise
		strategy imitation~\cite{traulsen:2006a}, these probabilities are
		given by
		\begin{equation}
		T^{\pm}(k)=\frac{k(Z-k)}{Z^2}\left( 1+e^{\mp\beta[\Pi_j(k)-\Pi_i(k)]} \right)^{-1},
		\end{equation}
		where $\beta$ represents the intensity of selection, i.e., the strength individuals base their decision to imitate the others.
	}
	
	In the case of neutral drift, the fixation probability among all pairs
	of strategies is $\rho_{ij}=\eta=1/Z$. The probabilities defined by
	Eq~(\ref{eq:rho}) determine a transition matrix of a Markov chain among
	strategies, assuming a sufficiently low mutation rate
	\cite{wu:2012}. The non-diagonal elements of this matrix are
	$\mathcal{T}_{ij}=\rho_{ij}\nu^{-1}$, and the diagonal is
	$\mathcal{T}_{ii}=1-\sum_j \mathcal{T}_{ij}$, where $\nu$ is the total number of
	strategies.  The normalised eigenvector associated with the first
	eigenvalue of that matrix provides the stationary distribution $D_i$
	of strategies~\cite{imhof:2005,fudenberg:2006}, which represents the
	relative time the population spends adopting strategy $i$.  The
	transition probability $\mathcal{G}_{g_ig_j}$ between two groups of strategies
	$g_i$ and $g_j$ is computed as:
	\begin{equation}
	\mathcal{G}_{g_ig_j}=\sum_{i\in g_i, j\in g_j} \mathcal{T}_{ij}\frac{D_i}{\sum_{k\in g_i}D_k}
	\end{equation}

	\subsection*{Drifting groups}
	\label{subsec:groups}
	
	In order to analyse the specific dynamics and success of the different
	strategies, we identify groups of strategies that show equivalent
	behaviour. Strategies [00\,$\ast$$\ast$$\alpha_1$$\alpha_2$]
	and [11\,$\alpha_1$$\alpha_2$$\ast$$\ast$]
	formed 8 groups (for each combination of $\alpha_i$, and varying
	$\ast$ within each group) whose strategies show neutral drift among
	them if the noise in signal is low enough. Specifically, when
	approximately $\epsilon<Q/N$ and $\epsilon<1-Q/N$ for the first and
	second set of groups, respectively, since the actual average number of
	individuals that signal in a group is $N_S-(2N_S-N)\epsilon$ when $N_S$
	of them intend to do it. In our analysis we assume $\epsilon=0.01$,
	therefore the previous groups are drifting groups (except for $Q=N$
	for the second set of groups). This does not mean that the
	probabilities of the strategies within a drifting group in the
	stationary distribution are the same, since, as we show, there usually
	exist some strategies that are able to invade others more
	successfully.
	
	Since the initial conditions of the iterated game assume that all
	individuals begin defecting, except because of noise, strategies
	[$\alpha_1$$\alpha_2$\,10$\alpha_3$$\alpha_4$]
	are equivalent to [$\alpha_1$$\alpha_2$\,00$\alpha_3$$\alpha_4$],
	and [$\alpha_1$$\alpha_2$\,$\alpha_3$$\alpha_4$10]
	to [$\alpha_1$$\alpha_2$\,$\alpha_3$$\alpha_4$00],
	if the noise is low and the threshold $M$ is high enough to prevent
	that the bit that differentiates them plays any role.  The probability
	that less than $M$ individuals commit an error acting due to noise in
	one round follows the binomial cumulative distribution function:
	\begin{equation}
	Pr_\epsilon(<M)=\sum_{k=0}^{M-1}\binom{N}{k}\epsilon^k (1-\epsilon)^{N-k}.
	\end{equation}  
	The probability that this happens in every round of the game (except
	the last one, in which case it would not have any effect) is
	$\left[Pr_\epsilon(<M) \right]^{\mathcal{R}-1}$, where $\mathcal{R}=(1-\omega)^{-1}$ is
	the average number of rounds of the game.
	Note that it would be possible that in a round $M$ is reached but in
	the next, again due to errors, individuals that would have started to
	cooperate defect again, correcting the first errors.
	This is the reason why, even when $\omega=1$ and all the states can be reached,  these drifting groups still appear, as we have seen in our analysis.

	\subsection*{Alternative scenarios}
	\label{subsec:self}
	
	Our model includes strategies that can decide their action taking into
	account if the number of individuals that cooperated in the past is
	higher or lower than the threshold $M$. We also analysed a scenario
	where---instead of using memory of past interactions---individuals base their decision on
	the actual reception of a benefit. We refer to this case as the
	\textit{benefit-perceived scenario}. The model remains the same,
	except when the ecological context is $\tilde{G}$. There exist two
	main possible interpretations of $\tilde{G}$: a scenario without
	resources or a scenario with abundance. In the former, individuals are
	not obtaining any benefit (equivalent in the regular scenario to
	consider that $M$ was not reached), whereas in the latter every
	individual obtains a benefit (equivalent to consider that $M$ is
	always reached). 
	
	A second alternative scenario is introduced to study the effects
		of self-awareness. So far, we have assumed that individuals can
	base their action on the number of individuals that cooperated within
	the group in the previous round (i.e., if the threshold $M$ was
	reached) without removing themselves from this counting. In the
	\textit{self-aware scenario}, we introduce another parameter
	$R\in[1,N-1]$, that stands for the number of individuals other than
	the focal player that cooperated in the previous round, and serves as
	a threshold for the individuals to decide their current action. As a
	consequence, $R$ can be different from $M$, which represents the
	number of individuals needed to obtain the benefit. The strategy space
	is modified in its acting module as $a_{\gamma,\gamma'}$, where now
	$\gamma'=\{R,\tilde{R}\}$.
	As a consequence, $\mathbf{A}$ also changes since $Pr^k(i,j)$
	corresponds now to the probability mass function of the Poisson
	binomial distribution~\cite{hong2013computing}, i.e., the probability
	that $N_{C,k}(j)$ $k$-individuals cooperate among the total $N_{k}$ in
	the group, considering that the probability of cooperation of each
	individual is $P_{C|C}^k(i)$ for $N_{C,k}(i)$ of them and
	$P_{C|D}^k(i)$ for the remaining $N_{k}(i)-N_{C,k}(i)$ ones, where
	\begin{equation}
	\begin{split}
	P_{C|C}^k(i)&=\Theta\left(N_S-Q\right)\Theta\left(N_C(i)-1-R\right)\,a_{Q,R}^k \\
	&+\Theta\left(N_S-Q\right)\left[1-\Theta\left(N_C(i)-1-R\right)\right]\,a_{Q,\tilde{R}}^k \\  
	&+\left[1-\Theta\left(N_S-Q\right)\right]\Theta\left(N_C(i)-1-R\right)\,a_{\tilde{Q},R}^k \\  
	&+\left[1-\Theta\left(N_S-Q\right)\right]\left[1-\Theta\left(N_C(i)-1-R\right)\right]\,a_{\tilde{Q},\tilde{R}}^k \\
	P_{C|D}^k(i)&=\Theta\left(N_S-Q\right)\Theta\left(N_C(i)-R\right)\,a_{Q,R}^k \\
	&+\Theta\left(N_S-Q\right)\left[1-\Theta\left(N_C(i)-R\right)\right]\,a_{Q,\tilde{R}}^k \\  
	&+\left[1-\Theta\left(N_S-Q\right)\right]\Theta\left(N_C(i)-R\right)\,a_{\tilde{Q},R}^k \\  
	&+\left[1-\Theta\left(N_S-Q\right)\right]\left[1-\Theta\left(N_C(i)-R\right)\right]\,a_{\tilde{Q},\tilde{R}}^k. 
	\end{split}
	\end{equation}

\section*{S2. Representative groups of strategies}

Among all the possible strategies that can use signalling and
reciprocity (condition \textbf{S+R}), we have identified the strategies
and drifting groups of strategies (i.e, groups in which strategies
show neutral drift among them) that are the most representative in the
stationary distributions for a wide range of parameters. Here we
describe these groups with more detail (see the main text for a
summary):

\begin{itemize}
\item \textit{Signalling to cooperate} (\textbf{SC}): strategy
  [10\,0011] signals under state $G$ and cooperates when the group signals. Strategies [10\,1011] joins the prototype in a drifting group if $M$ is high enough.
\item \textit{Signalling to defect} (\textbf{SD}): strategy [01\,1100]
  signals under state $\tilde{G}$ and
  only cooperates if the group is not signalling. \textbf{SD} is
  symmetric to \textbf{SC}, and in a similar way, strategy [10\,1110]
  forms a drifting group with it for high values of $M$.
\item \textit{Signalling to cooperate and opportunistic} (\textbf{SC-O}): strategy [10\,0001] is similar to the prototype of \textbf{SC}, but when it is time to cooperate, it only does it if $M$ was not reached previously.    
\item \textit{Signalling to defect and opportunistic} (\textbf{SD-O}): strategy [01\,0100] shows the same relation with \textbf{SD} as \textbf{SC-O} with \textbf{SC}. 
\item \textit{Free-riding in signalling but using the lack of signalling to cooperate} (\textbf{FR-C}): strategy [00\,1100] acts similarly as \textbf{SD} but does not signal under any circumstance. It constitutes a drifting group with [00\,11$\ast$$\ast$] strategies.
\item \textit{Free-riding in signalling and pure defecting} (\textbf{FR-D}): strategy [00\,0000] is the pure defector; it does not signal neither cooperates. It is part of the drifting group [00\,00$\ast$$\ast$].
\item \textit{Free-riding in signalling and opportunistic} (\textbf{FR-O})): strategy [00\,0100] behaves as \textbf{SD-O} without signalling. Its drifting group is [00\,01$\ast$$\ast$].
\item \textit{Free-riding in signalling and following} (\textbf{FR-F})): strategy [00\,1000] do not signal and cooperates only if the group was also not signalling and the $M$ was previously reached. Its drifting group is [00\,01$\ast$$\ast$]. This group also drifts with \textbf{FR-C} when $M$ is low, and with \textbf{FR-D} when $M$ is high enough. It only appears independently under the benefit-perceived scenario.

\end{itemize} 

One can see that all these important groups use the information from
signals to decide how to act (except the full defector
\textbf{FR-D}). However only some of them contribute actively
using signals under $G$ or $\tilde{G}$
(\textbf{SC}, \textbf{SD}, \textbf{SC-O}, and \textbf{SD-O}); the rest
just exploit signals for action, without paying the costs of
  signalling (\textbf{FR-C}, \textbf{FR-D}, \textbf{FR-O}, and
\textbf{FR-F}). Some of these strategies combine this information with
reciprocity (\textbf{SC-O}, \textbf{SD-O}, \textbf{FR-O} and
\textbf{FR-F}), but none of them use only reciprocity to make their
decisions.

\section*{S3. Influence of parameters of the model}

\paragraph{Emergence of signalling strategies.} For moderate $c_S$,
signalling is exploited and acquires a clear meaning, standing for the
identification of rare ecological conditions. This is clearly visible
in Fig.~2 of the main text for $M=5$ (see also
Fig.~\ref{fig:signal-action_M7} for $M=7$), which shows the prevalence
of all the signalling strategies within the $\{\lambda,c_S\}$
parameter space, while the dominating groups are displayed in Fig.~4
of the main text and also Fig.~\ref{fig:groupAGR_w09} for a finite
number of rounds ($\omega = 0.9$).
When $c_S$ is negligible, strategies that exploit signals to cooperate
(i.e., \sSC or \sSCO) and to defect (\sSD or \sSDO) dominate equally.
This is also visible in the invasion graphs from
Fig.~\ref{fig:graphs_S_l0.5_c0_M5} and
\ref{fig:graphs_S_l0.5_c0_M7}, in which we note also that the two
signalling strategies do not invade each other, as a result of the
different meaning associated to signals which prevents their
coexistence.
When $c_S$ takes a non-null but low value, \sSC and \sSCO prevail for
lower values of $\lambda$, while \sSD and \sSDO dominate for higher
ones, since it is less costly to signal for the less frequent
ecological condition. However this symmetry is quickly broken: whereas
\sSC and \sSCO continue dominating when $\lambda\leqslant 0.5$, \sSD
and \sSDO disappear quickly (see also the invasion graphs in Fig.~\ref{fig:graphs_S_l0.5_c0.1_M5} and \ref{fig:graphs_S_l0.3_c0.3_M7}).
Interestingly, unless $\lambda$ is close to an extreme, \sSC and
  \sSD never invade each other, neither do \sSCO and \sSDO (at least
  not with a probability higher than the neutral drift): the success
  of one or the other are the result of the interaction with other
  strategies, especially those that free-ride the signalling costs
  (see Fig.~\ref{fig:graphs_S}).
The asymmetric benefit for different environmental conditions result
in asymmetries in the strategies, so that signalling for the
unprofitable conditions results in lower advantage, and is therefore
subsumed by strategies that mostly cooperate but free ride the
signalling cost (see below). Given that coexistence of signals with
different meaning is not possible, different signalling strategies
mostly appear in isolation.

\paragraph{Free-riding on signalling costs.} When $c_S$ takes high
values, strategies that dominate do not signal under any circumstance
and only specify their actions, coherently, when the quorum in
signalling was not reached. Depending on the frequency of each
ecological condition, these actions may differ. When $\lambda$ is not
high, \sSC and \sSCO eventually become too expensive and \sFRD (full
defection) takes their place, since cooperation becomes a cost not
worthy to be paid for the few chances of obtaining a benefit in the
$G$ state (see also Fig.~\ref{fig:graphs_S_l0.5_c0.5_M5}). On the
other hand,  \sFRC prevails over \sSD
even for moderate values of $c_S$, because both employ the same
strategy to act (i.e., cooperate when there is no signal), but the
former does not pay any cost for signalling (see
Fig.~\ref{fig:graphs_S_l0.9_c0.3_M7}).
\sFRD cannot exploit \sSC in the same way because the latter
  cooperates only when the signalling quorum is attained, while in the
  absence of signalling, no benefit is obtained.  Hence, \sFRD cannot
  obtain any benefit by only free-riding the costs of signalling, and
  dominates only when $c_S$ is too high.
  Note that \sFRD hardly ever invades \sSC in
  Fig.~\ref{fig:graphs_S}. On the contrary, \sSC does invade \sFRD for
  moderate-low values of $c_S$, as discussed before.
In between \sFRC and \sFRD, there appears the opportunistic strategy
\sFRO. An individual playing \sFRO within a cooperative group acts as
a defector. Conversely, among defectors, \sFRO cooperates. It may look
like a counter-intuitive behaviour, but this strategy allows to
exploit cooperators, recognising when the own contribution is not
necessary to obtain the benefit, and if not enough individuals
cooperate, \sFRO makes an effort to sum up enough cooperators. This
elaborated behaviour has the objective to maximise the number of times
a benefit is obtained, but working for it as less as possible. When
individuals of an homogeneous population follow this strategy, they
will cooperate half of the time. This behaviour only pays when the $G$
state appears with a high frequency, otherwise defectors \sFRD take
over (see also Fig.~\ref{fig:graphs_S_l0.8_c0.3_M7} and
\ref{fig:graphs_S_l0.9_c0.5_M5}).
Overall, a general pattern among strategies that share the same
signalling behaviour can be found: \sFRO can invade \sFRC and \sFRD
can invade both, but never in the opposite direction (see
Fig.~\ref{fig:graphs_S}). The same happens with \sSCO invading \sSC,
and \sSDO invading \sSD. As observed in other studies based on public
good games, there exist an evolutionary tendency to defection when the
benefits-cost trade-off is not benign. Cooperative strategies become
successful when they can invade other strategies with a different
signalling behaviour, keeping defectors at bay without directly
invading defectors that share the same signalling behaviour.

\paragraph{Cooperation threshold and signalling quorum.} Thresholds $M$ and $Q$ have a significant role in the dynamics of the
system. Fig.~4 in the main text and Fig.~\ref{fig:groupAGR_w09} show how $M$ plays
a fundamental role. Cooperation is favoured with the increase of $M$:
cooperative strategies (\sSC, \sSD, and \sFRC) dominate over their
opportunistic counterparts (namely, \sSCO, \sSDO and \sFRO) when
$M > 5$, while full defection (\sFRD) dominates across large portions
of the parameter space for low values of $M$. Indeed, higher values of
$M$ impose stronger constraints for cooperation in order to receive a
benefit. In such conditions, \sSCO and \sSDO are not able to invade
\sSC and \sSD, respectively, since an unconditional cooperative
  strategy that is based on signalling only is more advantageous than
  the conditional counterpart.
The impact of the threshold $Q$ is less important. Intermediate values
of $Q$ are better for \sSC.
The
presence of \sSCO, on the other hand, increases with $Q$ when the
requirements for cooperation are mild ($3 \leq M\leq 7$).  This happens when the system has enough time to
self-organise, but not when the number of rounds is limited, as shown
in Fig.~\ref{fig:groupAGR_w09}.
Both \sFRC and \sFRO lose importance against defectors
playing \sFRD when $Q$ increases since it is more difficult that the
signalling quorum is reached when $Q$ is high.

As already suggested in the discussion above, the thresholds $M$ and
$Q$ are strongly related the probability of making errors with respect
to what dictated by the strategy. For instance, $M$ can be interpreted
as the effectiveness of cooperation against errors. An increase in the
error probability can be represented by an increase of the cooperation
threshold $M$, because individuals that cooperate may not effectively
contribute to reach the threshold due to frequent errors in their
actions. Under this interpretation, \sFRO is less successful because
higher cooperation thresholds/errors jeopardise the positive
contribution towards reaching the threshold, leading to less occasions
for opportunistic behaviour and overall lower payoff. In the same way,
when individuals commit errors in assessing the signalling behaviour,
reaching the signalling quorum $Q$ is more difficult, hence we can
interpret higher errors as a larger quorum. When $Q$ is low, the
signalling quorum may be reached due to errors even when in
$\tilde{G}$, leading \sSC and \sSCO to cooperate without obtaining any
benefit. 
With signalling, we can consider also another source of error, that
is, using the signal in the wrong ecological condition. This error
$\epsilon_S$ represents the inability of individuals to correctly
discriminate $G$ from $\tilde{G}$, so that a signal is produced
  with probability
$(1-\epsilon_S)s_{\alpha}+\epsilon_Ss_{\bar{\alpha}}$. This
type of error could undermine the success of \sSC or
\sSD, which trust in the correct assessment of the ecological
  context to decide whether or not to cooperate. However, results
show that the signalling strategies are rather robust against this
kind of error, thanks to the aggregation of information from
  multiple individuals and to the quorum mechanism. As a matter of
  fact, a rather high value of $\epsilon_S$ is necessary to
observe some effect, as can be seen in Fig.~\ref{fig:groupAGR_eps-w}.

\section*{S4. Invasibility preference}

The evolutionary process leading to the emergence of one or the
  other strategy is very complex, as shown by the invasibility graphs
  in Fig.~\ref{fig:graphs_S}, because every strategy plays a role, even
  when its probability in the stationary distribution is low. In spite of
this, it is possible to deduce the conditions under which a given
strategy is more likely to replace another strategy than \textit{vice
  versa}. Due to the complexity of the strategies themselves, we focus
only on the conditions on $c_S$ and $\lambda$ for those pairs of
strategies where both members emerge for the same combination of the
other parameters. The preference on the direction of the invasibility
between two strategies does not guarantee that the preferred strategy
has higher probability in the stationary distribution due to the
influence of the rest of the strategies. Therefore this analysis
should be taken as a way to corroborate the results that were already
discussed.

In the limit of large population size $Z$, strategy $A$ replaces $B$ in the population if
\begin{equation}
\sum_{N_A=1}^{N} W_A(N_A,N_B) > \sum_{N_A=0}^{N-1} W_B(N_A, N_B),
\end{equation}
where $W_\Lambda(N_A,N_B)$ is the payoff that a focal player following strategy $\Lambda$ obtains in a group of size $N$ formed by $N_A$ individuals playing strategy $A$ and $N-N_B$ playing strategy $B$~\cite{kurokawa2009emergence}. We apply this condition for some significant pairs of strategies $(A, B)$ assuming that $\epsilon\rightarrow 0$, $M,Q=\{1,\dots,N-1\}$ are integers (the closest greater integer is taken when the original parameter was not an integer), $\omega=1$, and $c=1$, obtaining the following conditions:

\begin{itemize}
	\item \textbf{(SC, FR-D)}:\mbox{    } $c_S< \left(\dfrac{r}{N}-1\right) +\left(\dfrac{Q-1}{N}\right)$
	\item \textbf{(SC-O, FR-D)}:\mbox{    } $c_S< \dfrac{1}{2} \left[ \left(\dfrac{r}{N}-1\right) +\left(\dfrac{Q-1}{N}\right)-\dfrac{\max\left\{0,M-Q\right\}}{N}\right]$
	\item \textbf{(FR-C, FR-D)}:\mbox{    } $\lambda>\dfrac{N}{r}$
	\item \textbf{(FR-O, FR-D)}:\mbox{    } $\lambda>\dfrac{N+M-1}{r}$
	\item \textbf{(SC, SC-O)} and \textbf{(SD, SD-O)}:\mbox{    }$r-2N+M>0$
	\item \textbf{(SC, FR-C)}:\mbox{    } $\dfrac{1}{\lambda}-c_S>\dfrac{2(N-Q)+1}{N}$
	\item \textbf{(SD, FR-C)}:\mbox{    } $c_S<\dfrac{1}{N}$
	\item \textbf{(FR-O, FR-C)}:\mbox{    } $\lambda<\dfrac{2N-M}{r}$
	\item \textbf{(SC, SD)}:\mbox{    } $\left(\lambda-\dfrac{1}{2}\right)c_S<\dfrac{2Q-N-1}{2N}$
\end{itemize}
These conditions generally match the qualitative behaviour observed in
Fig.~\ref{fig:graphs_S} and discussed in Section S3. Quantitative
  differences, when present, are justified by the approximations taken and by the
  fact that this analysis considers only single strategies---chosen as
  the prototype for each drifting group (see Section S2)---while the
  invasion graphs in Fig.~\ref{fig:graphs_S} take the entire drifting
  group into account.

\clearpage

\begin{table}
	\begin{center}
		\begin{tabular}{c|c|c|c|c|}
			\hhline{~----} 
			& Ecological conditions  & Signalling quorum & Memory on cooperation  & Number of \\
			&  $G$/$\tilde{G}$  & $Q$/$\tilde{Q}$  & $M$/$\tilde{M}$ & strategies \\
			\hhline{-====}
			\multicolumn{1}{|c|}{\sB} & - & - & - & $2$  \\
			\multicolumn{1}{|c|}{\sBp} & $a$ & - & - & $2^2$   \\
			\hhline{=====}
			\multicolumn{1}{|c|}{\sR} & - & - & $a$ & $2^2$ \\
			\multicolumn{1}{|c|}{\sRp} & $a$ & - & $a$ & $2^4$   \\
			\hhline{=====}
			\multicolumn{1}{|c|}{\sS} & $s$ & $a$ & - & $2^4$  \\
			\multicolumn{1}{|c|}{\sSp} & $s$, $a$ & $a$ & - & $2^6$  \\
			\hhline{=====}
			\multicolumn{1}{|c|}{\multirow{2}{*}{\sSR}} & \multirow{2}{*}{$s$} & \multirow{2}{*}{$a$} & \multirow{2}{*}{$a$} & \multirow{2}{*}{$2^6$}  \\
			\multicolumn{1}{|c|}{} & & & & \\
			\hline   
		\end{tabular}
	\end{center}
	\caption{\textbf{Summary of the combination of mechanisms
              analysed}. For each combination of mechanism exploited, we provide the possible input variables (ecological condition
            $G$/$\tilde{G}$, attainment of the signalling quorum
            $Q$/$\tilde{Q}$ and achievement of the cooperation
            threshold $M$/$\tilde{M}$), which determine the
            possibility to signal ($s$) and/or to act ($a$). The total
            number of resulting strategies for each experimental
            condition is also shown.} 
	\label{tab:summ}
\end{table}

\begin{figure}
  \centerline{\includegraphics[width=12cm]{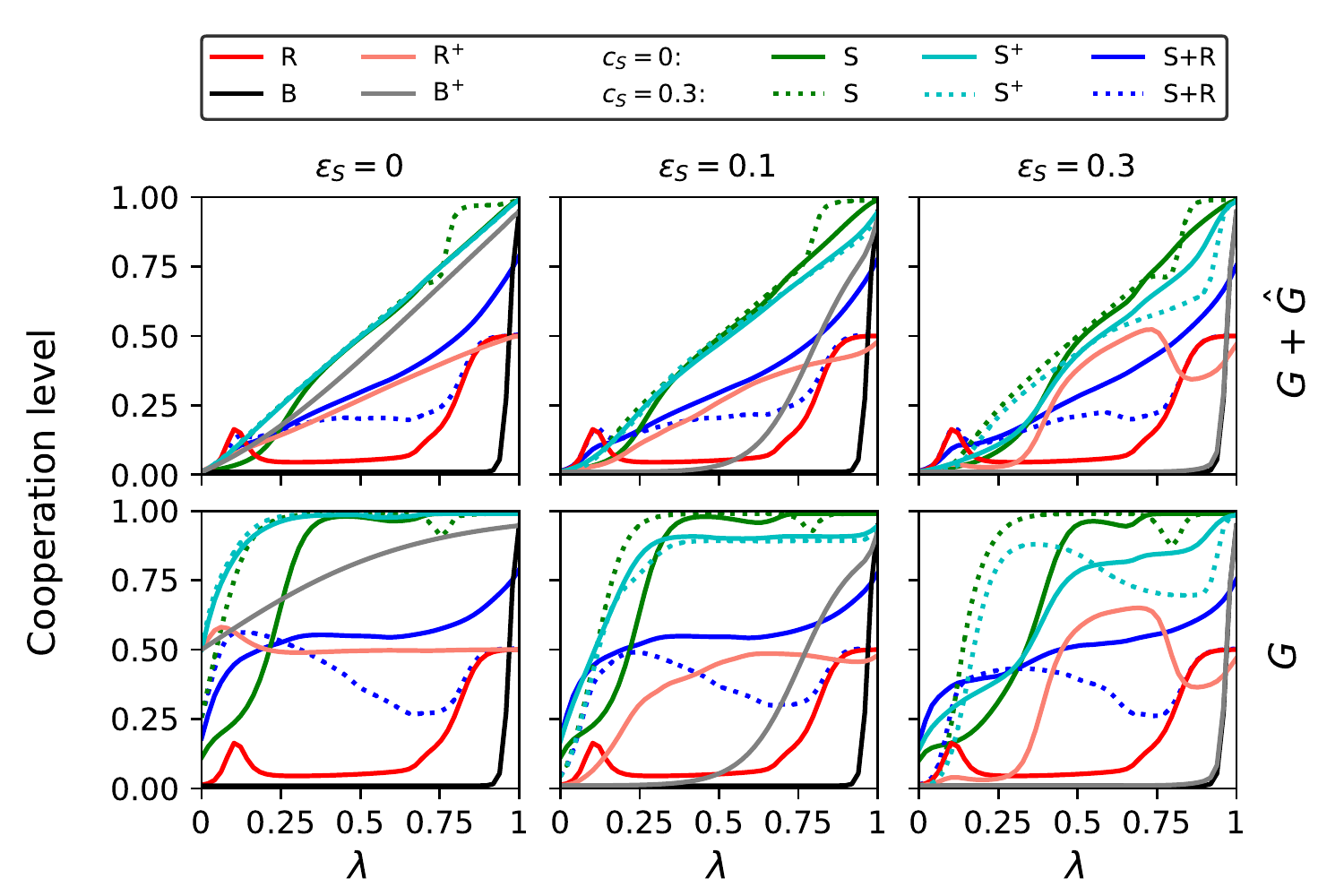}}
      \caption{ \textbf{Cooperation level attained under different
          evolutionary conditions.} Similarly to Fig.~1 in the main
        text, we consider reciprocity
        		(\sR), signalling (\sS), the combination of
        		them (\sSR), the baseline scenario (B), and strategies with the ability to discriminate between $G$ and $\tilde{G}$ (\sRp, \sSp, and \sBp, respectively). Different
        		values of the cost of signalling are
        		considered. The first row displays the level
        		    of cooperation attained when both states $G$ and $\tilde{G}$ are
        		    considered, and corresponds to Fig.~1 in the main text. The
        		    second row shows the level of cooperation when only the $G$
        		    state is considered. Model parameters: $M=5$, $Q=N/2$, $\omega=1$,
        		$\beta=1$, $r=10$, $c=1$, $\epsilon=0.01$, $N=9$, and $Z=100$.}      		
  \label{fig:coop}
\end{figure} 

\begin{figure}
  \centerline{\includegraphics[width=12cm]{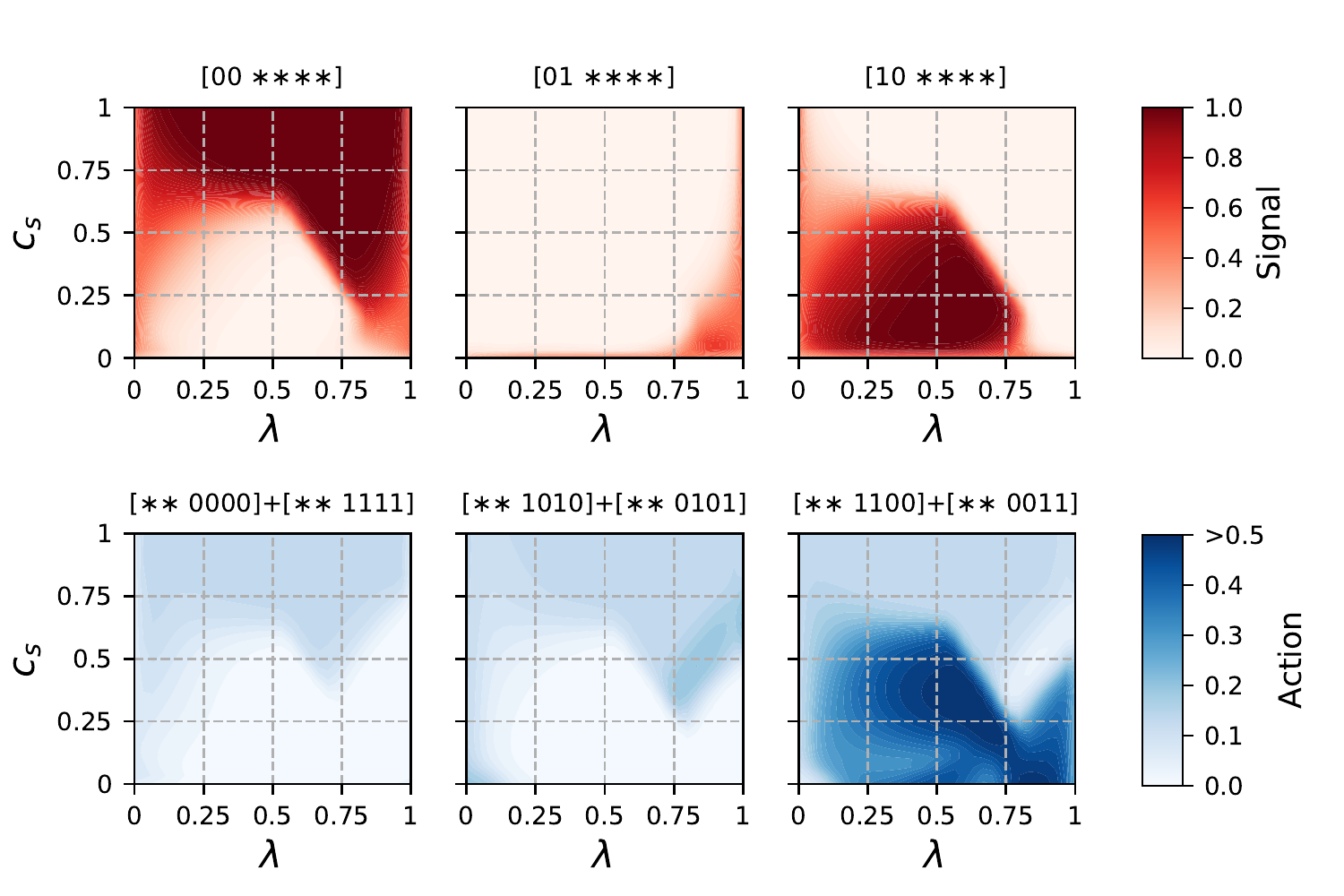}}
  \caption{\textbf{Prevalence of strategies grouped by signalling or
      acting behaviour.} Similarly to Fig.~2 in the main text, the
    plots show the aggregated probabilities in the stationary
    distribution of strategies across the $\{\lambda,c_S\}$ parameter
    space. Here, the behaviour for $M=7$ is displayed. Signalling
    strategies (top row) are grouped by the first two bits
    ($s_{G},s_{\tilde{G}}$), ignoring the always-signalling group
    which has negligible prevalence. Concerning the action part
    (bottom row), we show groups of strategies that exploit either
    reciprocity (middle panel) or signalling (right panel), in
    comparison to strategies that do not use any mechanism (left
    panel).  Model parameters: $Q=N/2$, $\omega=1$, $\beta=1$, $r=10$,
    $c=1$, $\epsilon=0.01$, $N=9$, and $Z=100$.}
  \label{fig:signal-action_M7}
\end{figure}

\begin{figure}
  \centerline{\includegraphics[width=17cm]{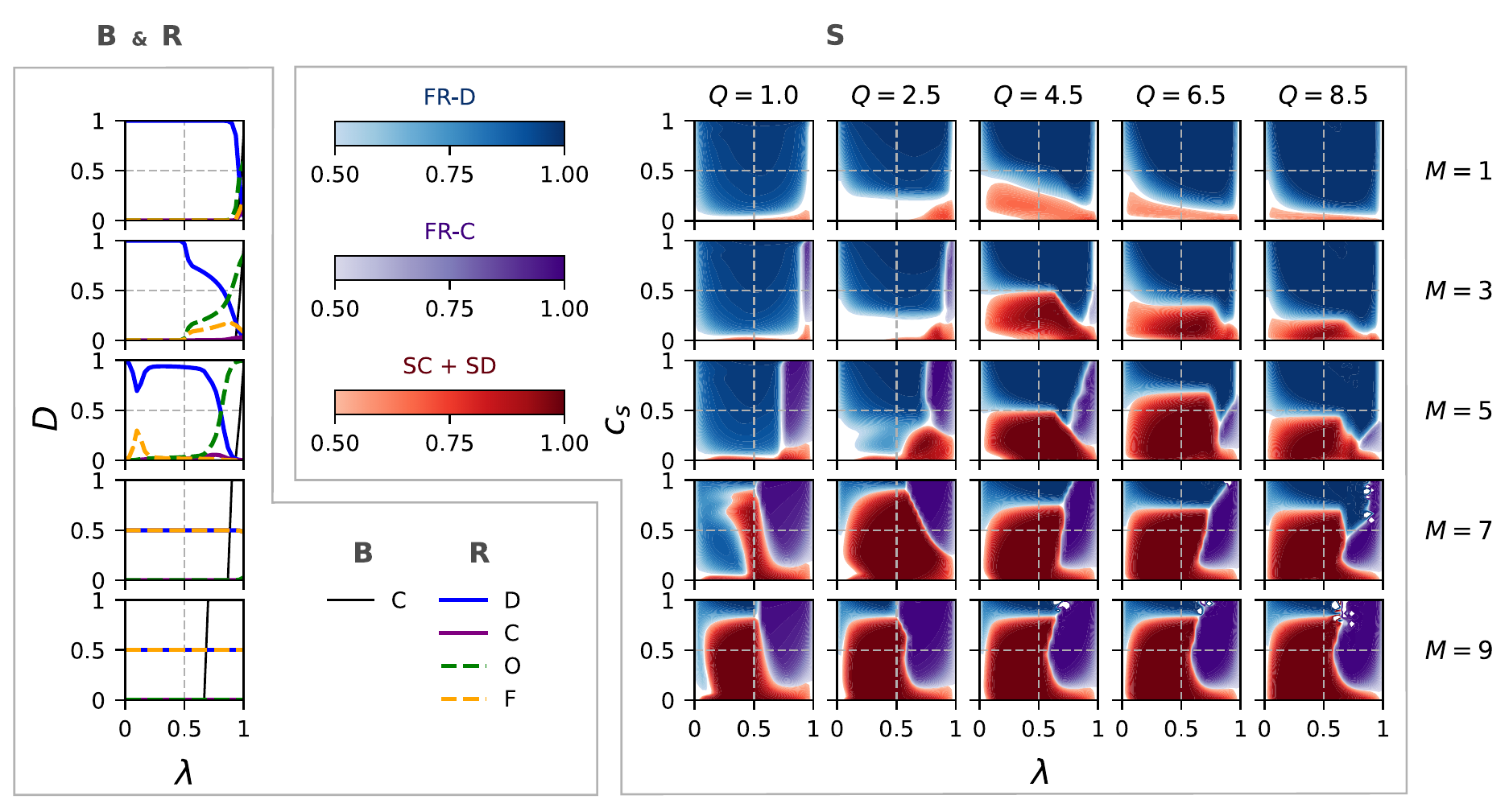}}
  \caption{\textbf{Stationary distribution in the \textbf{B}, \textbf{R}
      and \textbf{S} condition}. The baseline \textbf{B} and the
    reciprocity \textbf{R} conditions are displayed in the leftmost
    column, and only signalling \textbf{S} is shown with varying
    quorum $Q$ in the remaining panels. For \textbf{B} and \textbf{R},
    the column shows the probabilities of all the possible strategies
    for different thresholds $M$. The other
    panels show the stationary distribution of the most successful
    group of strategies when signalling is the sole mechanism
    available (condition \textbf{S}), and for different values of $M$
    and $Q$. Only probabilities higher than $0.5$ have been displayed
    to facilitate the visualisation in overlapping areas. Model
    parameters: $\omega=1$, $\beta=1$, $r=10$, $c=1$, $\epsilon=0.01$,
    $N=9$, and $Z=100$.}
  \label{fig:SD_diffmech} 
\end{figure}

\begin{table}
  \begin{center}
    \begin{tabular}{cc|cc|cc|cc|}
      \hhline{~~------} 
      &  & \multicolumn{2}{|c|}{$\lambda=0.2$} & \multicolumn{2}{c|}{$\lambda=0.5$} & \multicolumn{2}{c|}{$\lambda=0.8$} \\
      \hhline{~-======}
    \multicolumn{1}{c|}{\Gape[0.5cm][0.25cm]{}} &  \sB & \sD & $1.0$ & \sD & $1.0$ & \sD & $1.0$ \\
		\cline{2-8}
	\multicolumn{1}{c|}{} & \multirow{2}{*}{\sBp}  & (\sC, \sD) & $0.65$  & (\sC, \sD) & $0.82$  & (\sC, \sD) & $0.92$ \\
	 \multicolumn{1}{c|}{}   &    & (\sD, \sD) & $0.2$  & (\sD, \sD) & $0.35$  & (\sD, \sD) & $0.08$ \\
      \hhline{~=======}
    \multicolumn{1}{c|}{} & \multirow{2}{*}{\sR} & \multirow{2}{*}{\sD} & \multirow{2}{*}{$0.92$} & \multirow{2}{*}{\sD} & \multirow{2}{*}{$0.93$} & \sD & $0.52$ \\
     \multicolumn{1}{c|}{} 	& 					&  &  &  &  & \sO & $0.42$ \\
		\cline{2-8}
	\multicolumn{1}{c|}{} & \multirow{3}{*}{\sRp}& (\sO, \sD) & $0.55$  &  \multirow{3}{*}{(\sO, \sD)} &  \multirow{3}{*}{$0.83$}  &  \multirow{2}{*}{(\sO, \sD)} &  \multirow{2}{*}{$0.76$} \\
	\multicolumn{1}{c|}{}   &     & (\sC, \sD) & $0.21$  & &   &  \multirow{2}{*}{(\sO, \sF)} &  \multirow{2}{*}{$0.18$} \\
	\multicolumn{1}{c|}{}   &     & (\sD, \sD) & $0.21$  &  &   &  &  \\
      \hhline{-=======}
      \multicolumn{1}{|c|}{\multirow{4}{*}{\rotatebox[origin=c]{90}{\small$c_S=0$}}} & \multirow{2}{*}{\sS} & \sSCSD & $0.44$ &  \multirow{2}{*}{\sSCSD} & \multirow{2}{*}{$0.98$} & \multirow{2}{*}{\sSCSD} & \multirow{2}{*}{$0.98$} \\
    	\multicolumn{1}{|c|}{}	& 			& \sD & $0.54$ &  & &  &  \\
		\cline{2-8}
	\multicolumn{1}{|c|}{\Gape[0.5cm][0.25cm]{}} & \sSp & (\sSCSD, \sD) & $0.60$ & (\sSCSD, \sD) & $0.64$ & (\sSCSD, \sD) & $0.64$ \\
      \hline \hline
    \multicolumn{1}{|c|}{\Gape[0.5cm][0.25cm]{} \multirow{2}{*}{\rotatebox[origin=c]{90}{\small$c_S=0.3$}}} & \sS & \sSC & $0.96$ & \sSC & $1.0$ & \sFRC & $0.79$ \\
		\cline{2-8}
	\multicolumn{1}{|c|}{\Gape[0.5cm][0.25cm]{}} & \sSp & (\sFRC, \sFRD) & $0.92$ & (\sFRC, \sFRD) & $0.97$ & (\sFRC, \sFRD) & $0.98$ \\
		\hline       
	       
    \end{tabular}
  \end{center}
  \caption{\textbf{Stationary distribution under every mechanism
        separately}. Stationary distribution of the main strategies or
      group of strategies that emerge under conditions \sB, \sBp, \sR, \sRp, \sS,
      and \sSp for different values of $\lambda$ assuming
      that there is no error in perceiving the ecological
        context ($\epsilon_S=0$). Strategies under \sBp, \sRp,
      and \sSp are shown as the composition of two strategies between
      parenthesis; individual chooses the first strategy when under
      $G$ and the second when under $\tilde{G}$. Numbers on the right
      of each strategy indicate their probability in
      the stationary distribution. Parameters of the model: $M=5$,
      $Q=N/2$, $\omega=1$, $\beta=1$, $r=10$, $c=1$, $\epsilon=0.01$, $N=9$, and $Z=100$.}
\label{tab:SD}
\end{table}

	\begin{figure}
		\centering	 
		\vspace{-0.4cm}
		\subcaptionbox{\textbf{$\lambda=0.5$, $c_S=0$, $M=5$}\label{fig:graphs_S_l0.5_c0_M5}}{
			\vspace{-0.6cm}\includegraphics[width=0.35\textwidth]{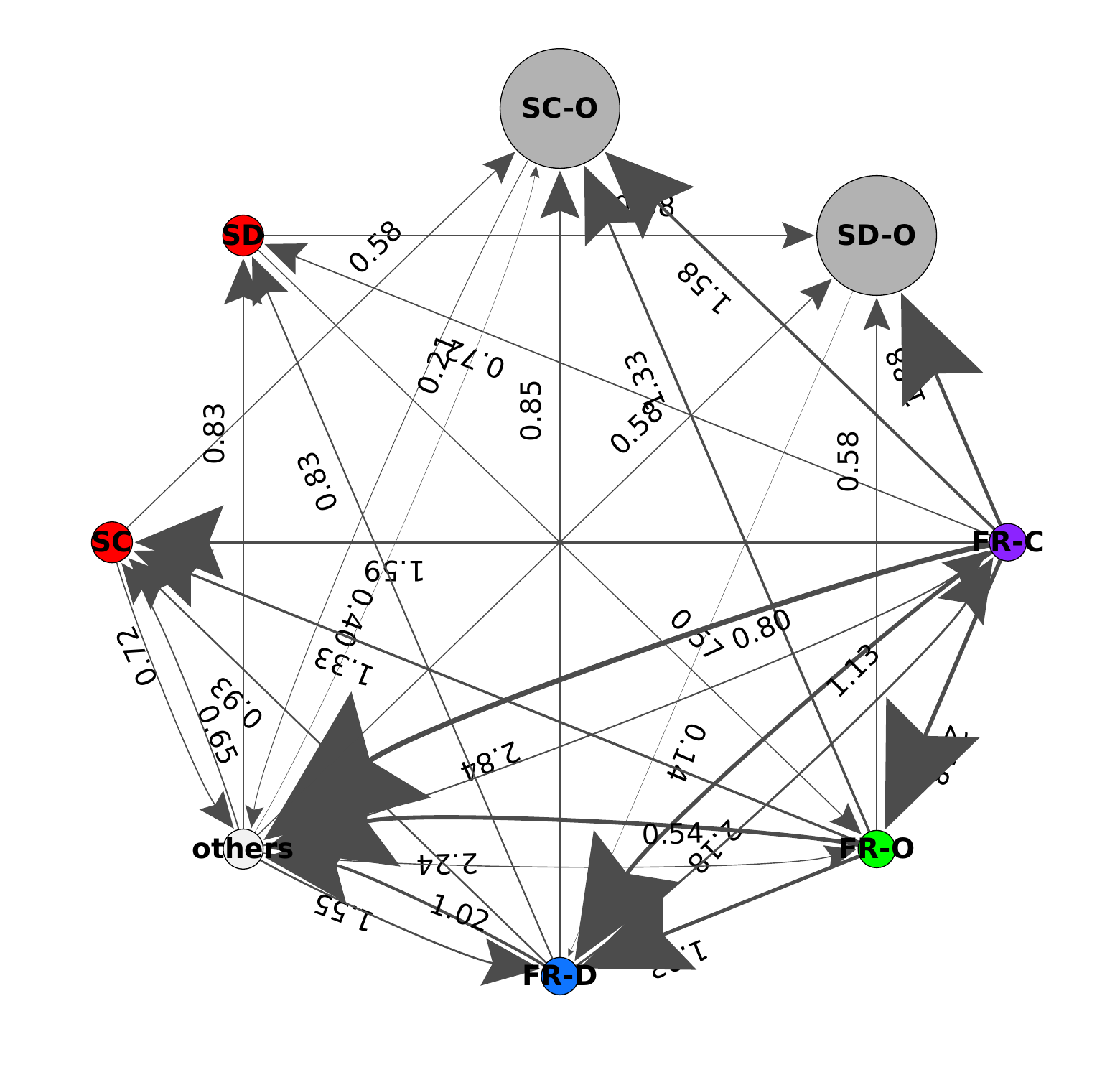}
		}   
		\vspace{0.3cm}
		\subcaptionbox{\textbf{$\lambda=0.5$, $c_S=0$, $M=7$}\label{fig:graphs_S_l0.5_c0_M7}}{
			\vspace{-0.6cm}\includegraphics[width=0.35\textwidth]{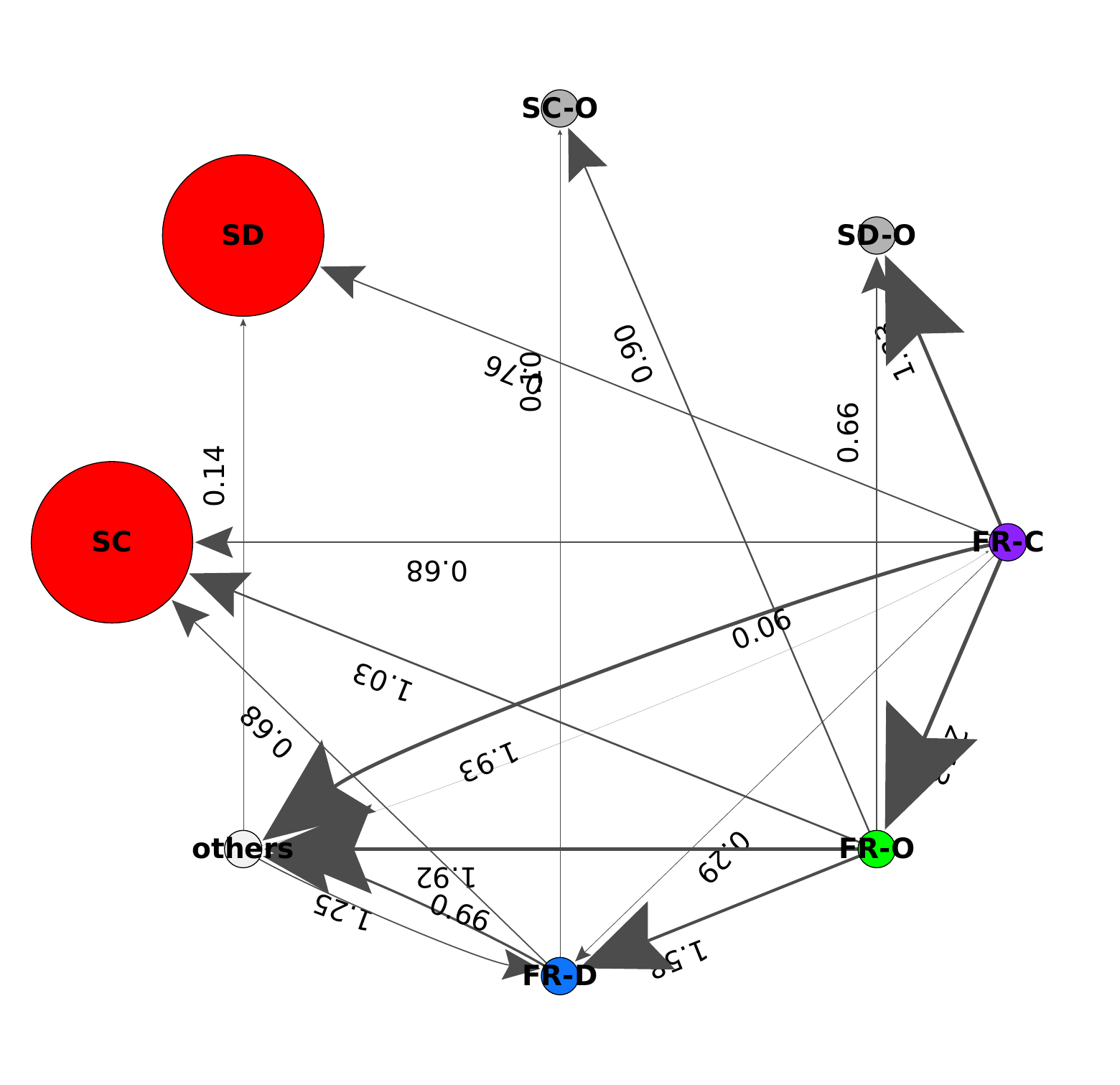}
		}  
		\subcaptionbox{\textbf{$\lambda=0.5$, $c_S=0.1$, $M=5$}\label{fig:graphs_S_l0.5_c0.1_M5}}{
			\vspace{-0.6cm}\includegraphics[width=0.35\textwidth]{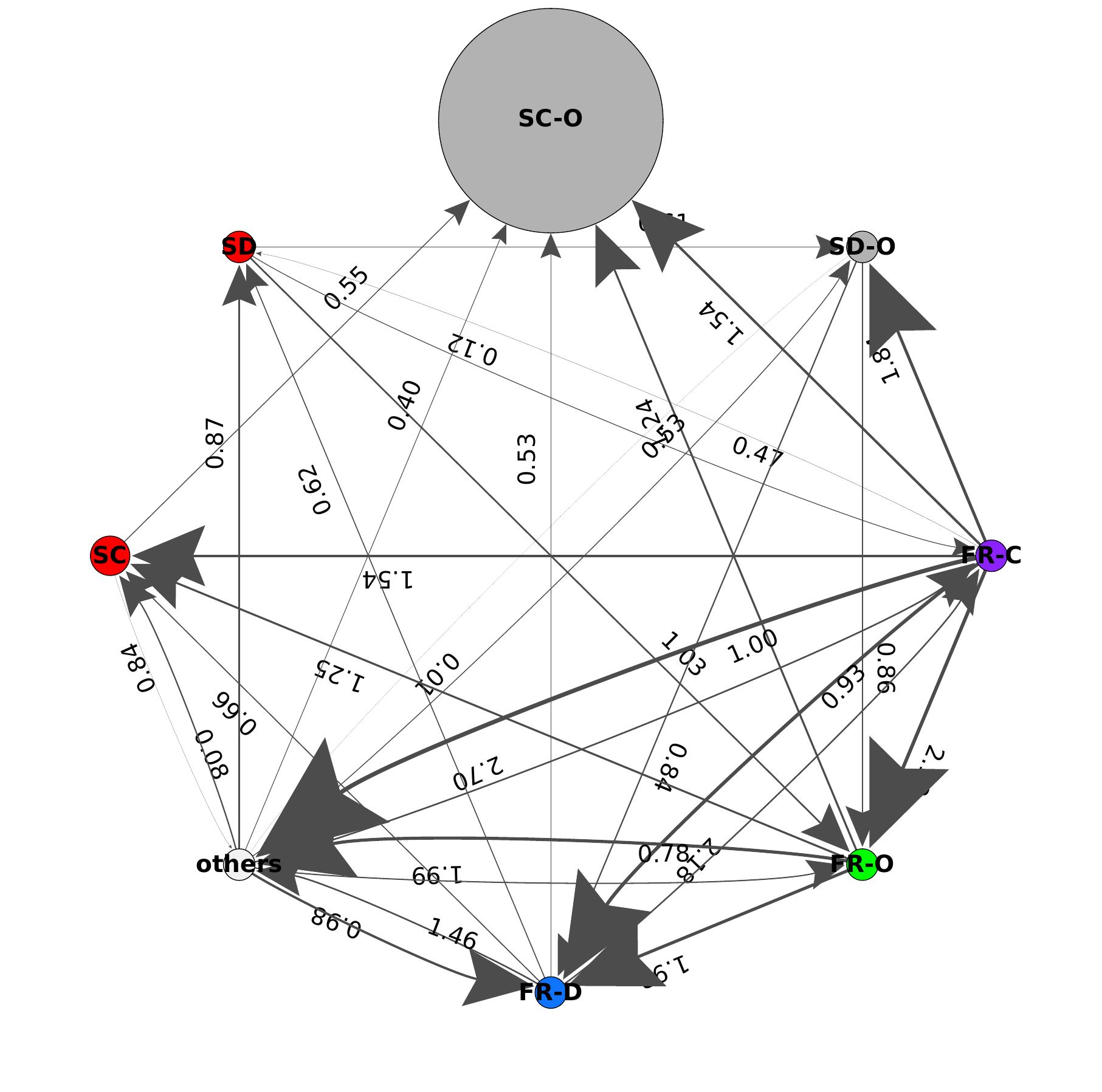}
		}
		\subcaptionbox{\textbf{$\lambda=0.3$, $c_S=0.3$, $M=7$}\label{fig:graphs_S_l0.3_c0.3_M7}}{
			\vspace{-0.6cm}\includegraphics[width=0.35\textwidth]{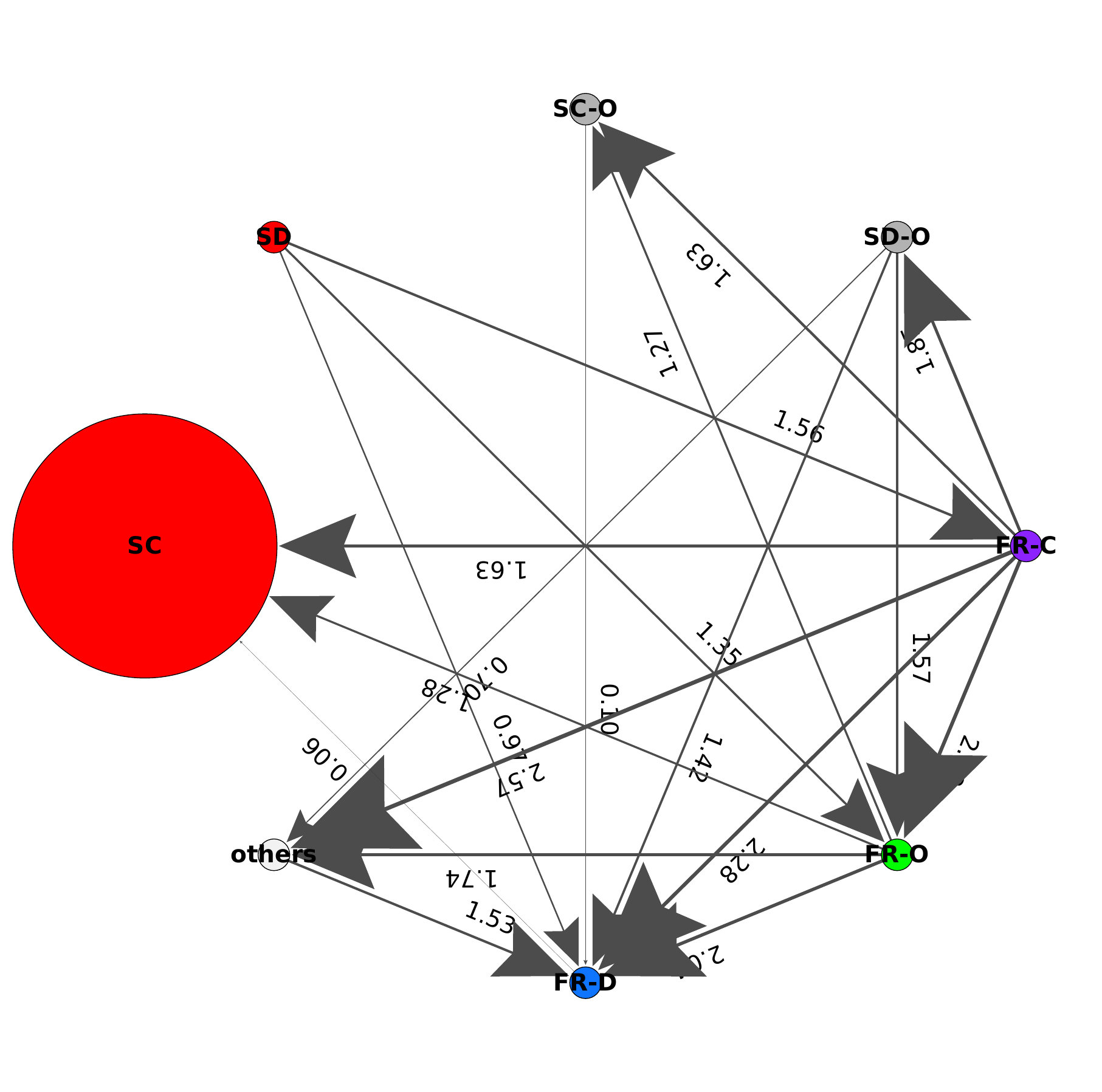}
		}
		\subcaptionbox{\textbf{$\lambda=0.5$, $c_S=0.5$, $M=5$}\label{fig:graphs_S_l0.5_c0.5_M5}}{
			\vspace{-0.2cm}\includegraphics[width=0.35\textwidth]{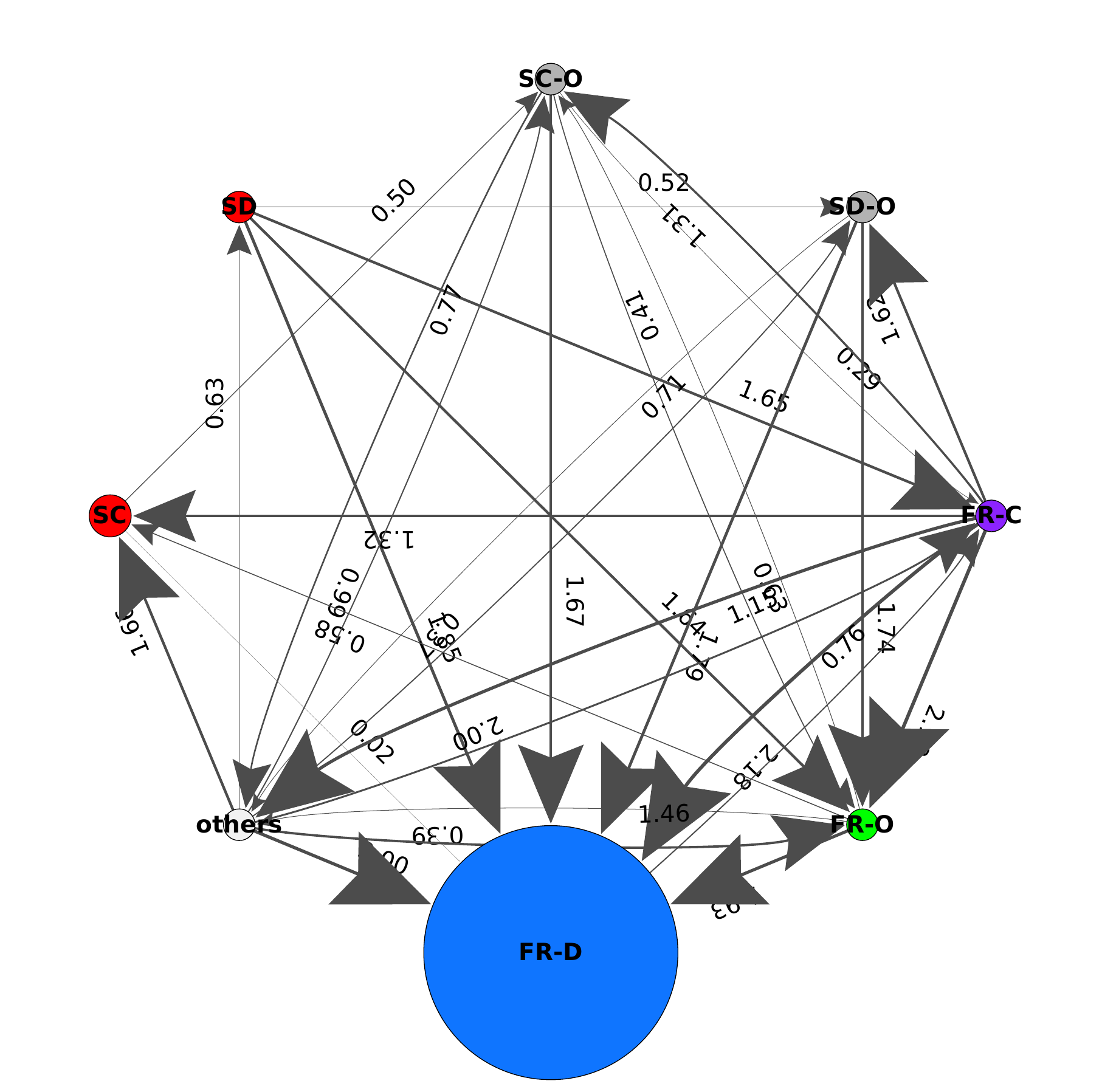}
		}
		\subcaptionbox{\textbf{$\lambda=0.8$, $c_S=0.3$, $M=7$}\label{fig:graphs_S_l0.8_c0.3_M7}}{
			\vspace{-0.2cm}\includegraphics[width=0.35\textwidth]{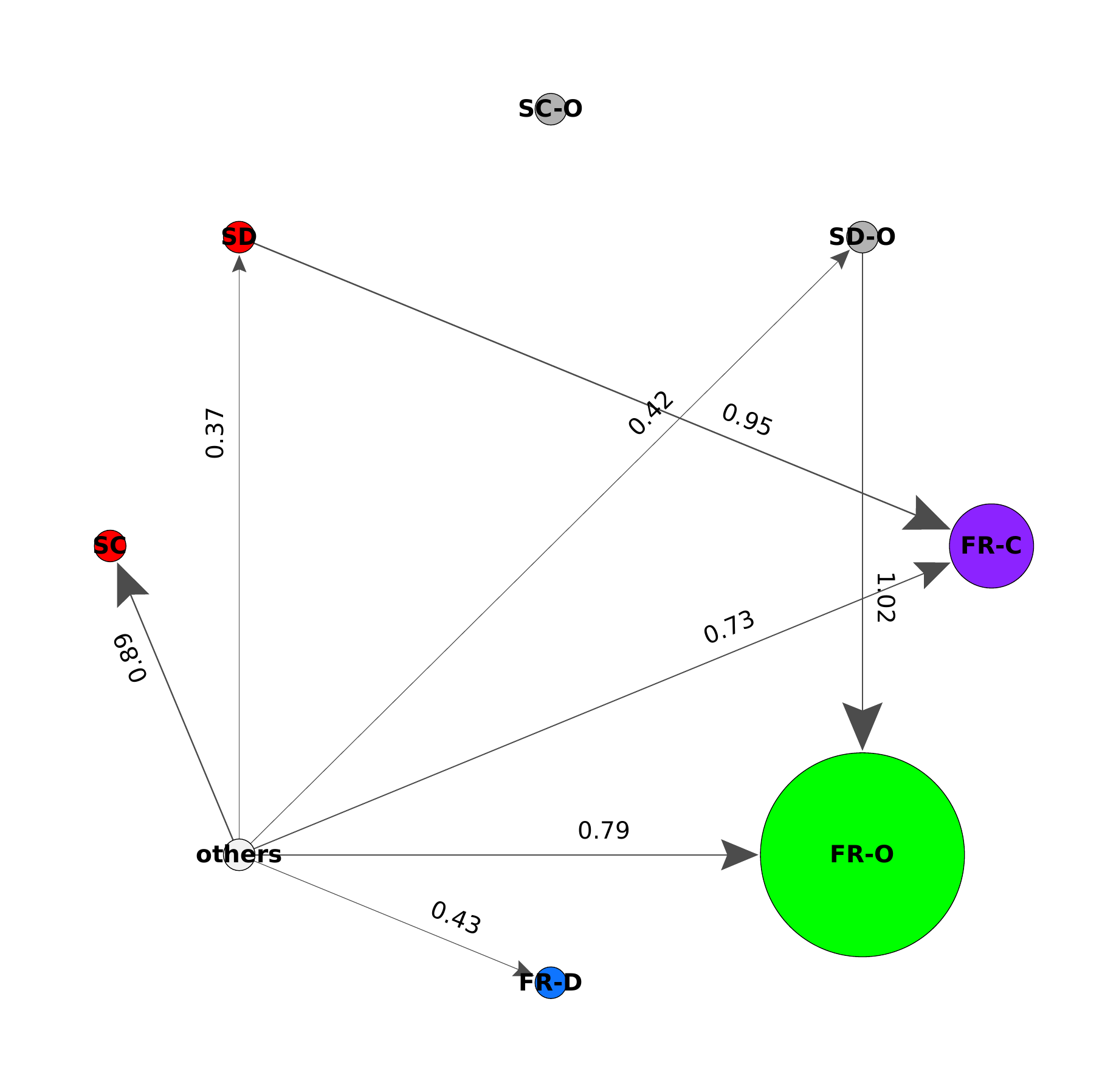}
		}
		\subcaptionbox{\textbf{$\lambda=0.9$, $c_S=0.5$, $M=5$}\label{fig:graphs_S_l0.9_c0.5_M5}}{ 
			\vspace{-0.6cm}\includegraphics[width=0.35\textwidth]{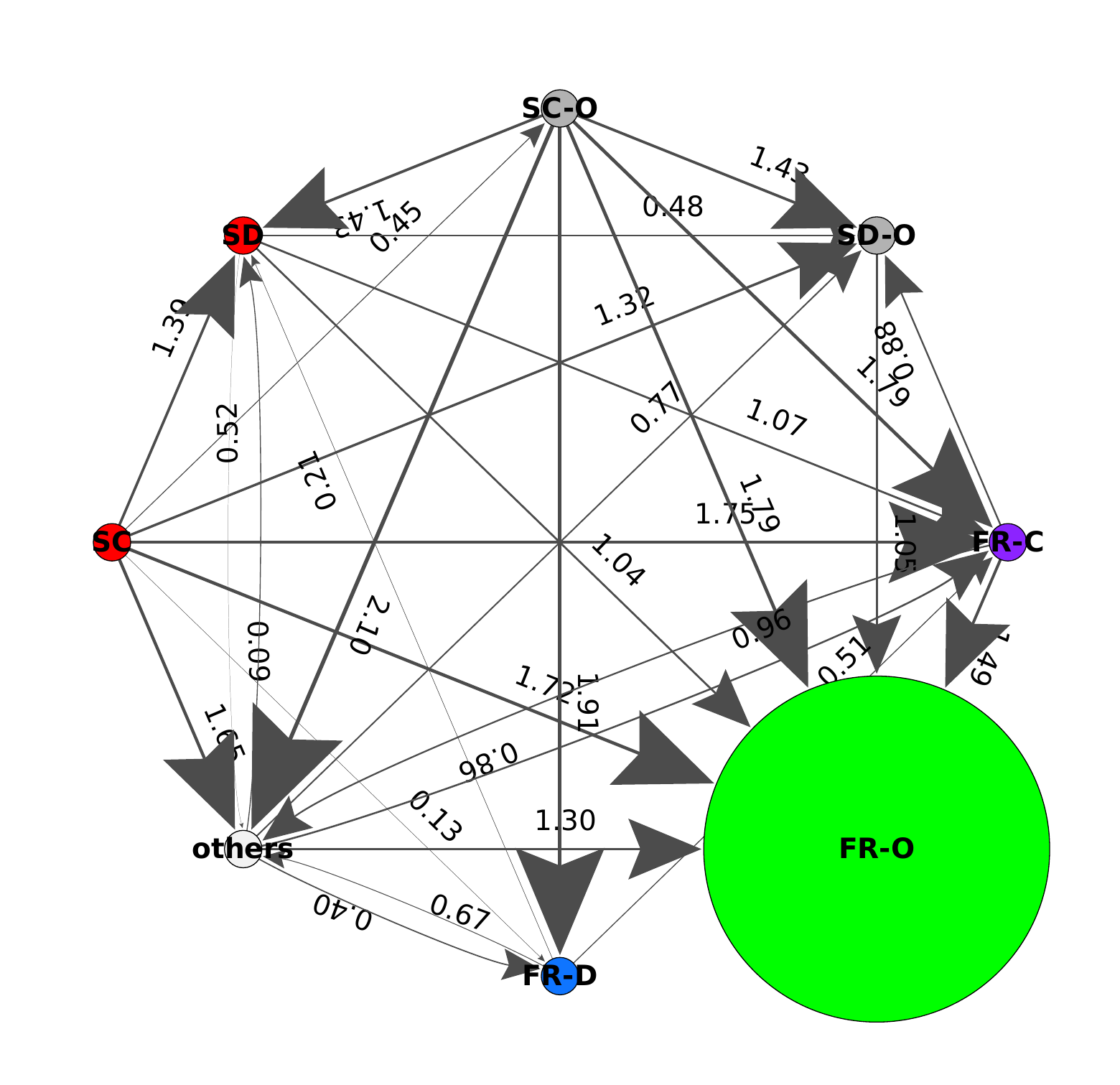}
		}
		\subcaptionbox{\textbf{$\lambda=0.9$, $c_S=0.3$, $M=7$}\label{fig:graphs_S_l0.9_c0.3_M7}}{
			\vspace{-0.6cm}\includegraphics[width=0.35\textwidth]{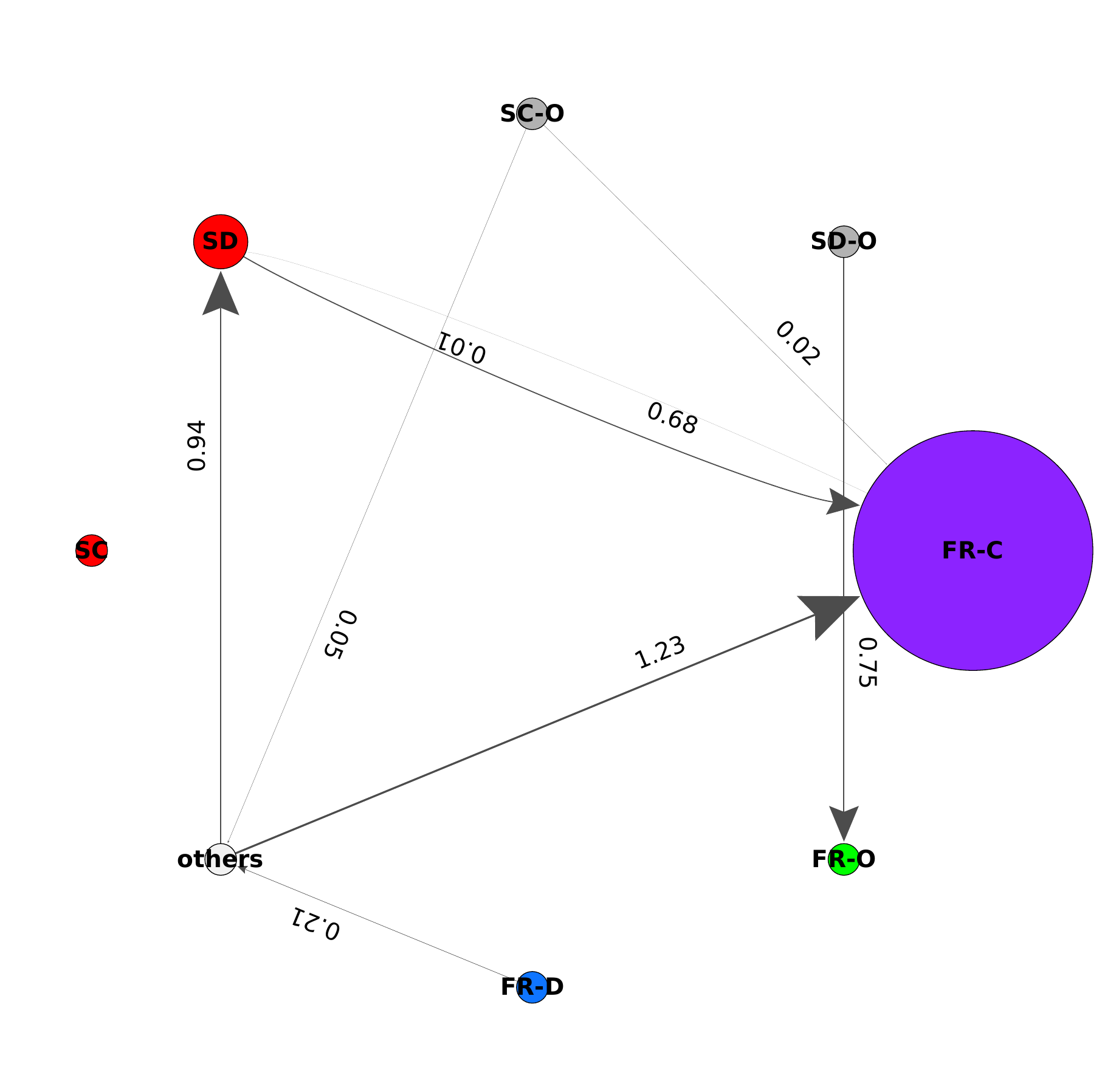}
		}
		\caption{\textbf{Graphs of invasion among the most
                    representatives groups of strategies}. Each panel
                  shows the invasion graph for different combinations
                  of $\lambda$, $c_S$ and $M$. The node size is
                  proportional to probabilities in the stationary
                  distribution. The arrow size and numbers above them
                  correspond to the logarithm of the normalised
                  transition probability among groups of strategies
                  $\boldsymbol{\mathcal{G}}(\nu\eta)^{-1}$. Parameters of the model: $Q=N/2$, $\omega=1$, $\beta=1$, $r=10$, $c=1$, $\epsilon=0.01$, $N=9$, and $Z=100$.}
		\label{fig:graphs_S}
	\end{figure}

	\begin{figure}
		\centerline{\includegraphics[width=12cm]{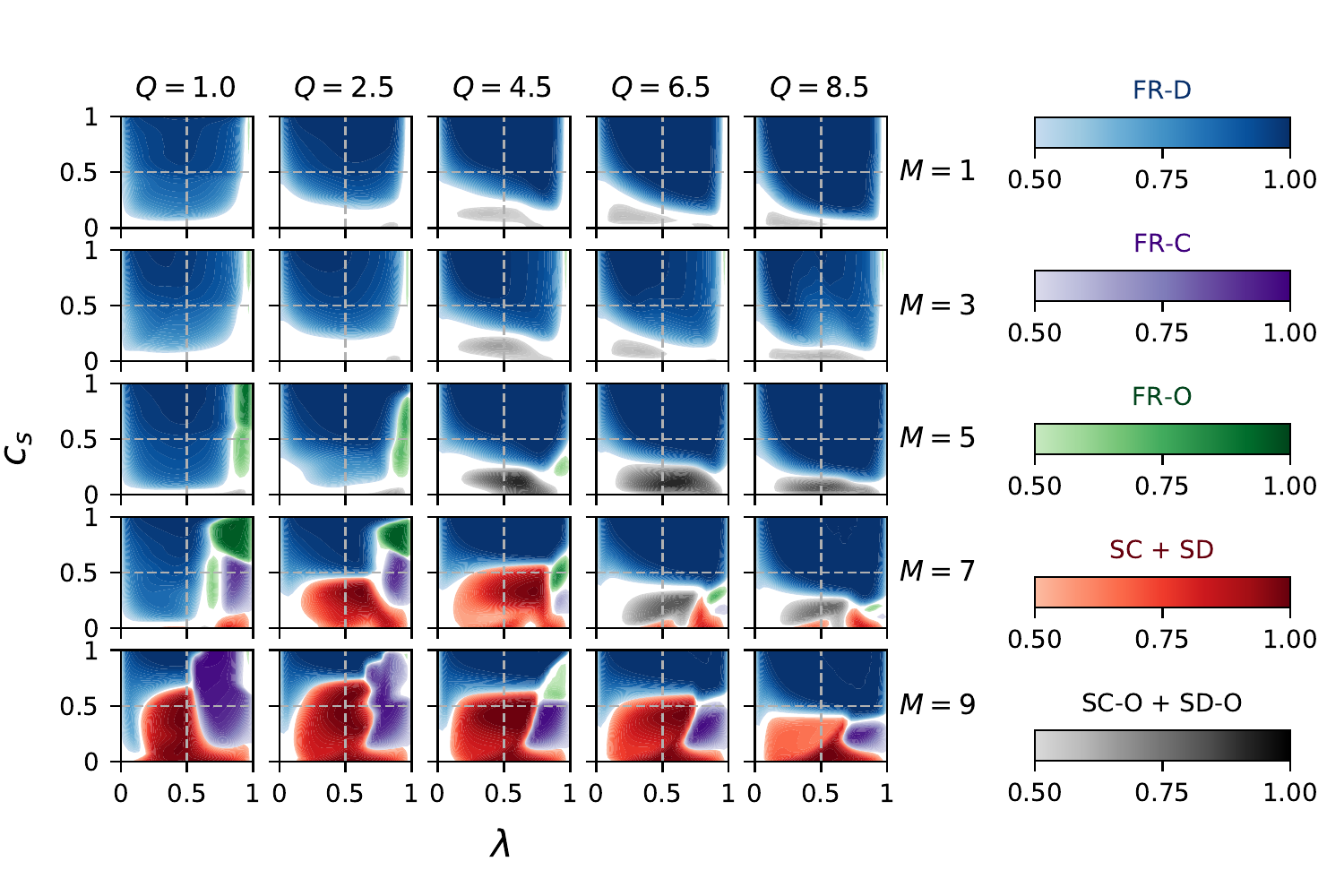}}
		\caption{\textbf{Stationary distribution of the main
                    groups of strategies  for $\omega=0.9$}.  Note that reducing the number of rounds
                  does not change in general the results. The main
                  difference with respect to Fig.~ 4 in the main text
                  takes place when $Q$ is high and $M$
                  intermediate-low. In this case, one can detect the
                  reduction of the success of \textbf{SC-O} and
                  \textbf{FR-O} strategies for a high cost of
                  signalling $c_S$. Model parameters: $\beta=1$,
                  $r=10$, $c=1$, $\epsilon=0.01$, $N=9$, and
                  $Z=100$.}
		\label{fig:groupAGR_w09}
	\end{figure}

	\begin{figure}
		\centerline{\includegraphics[width=12cm]{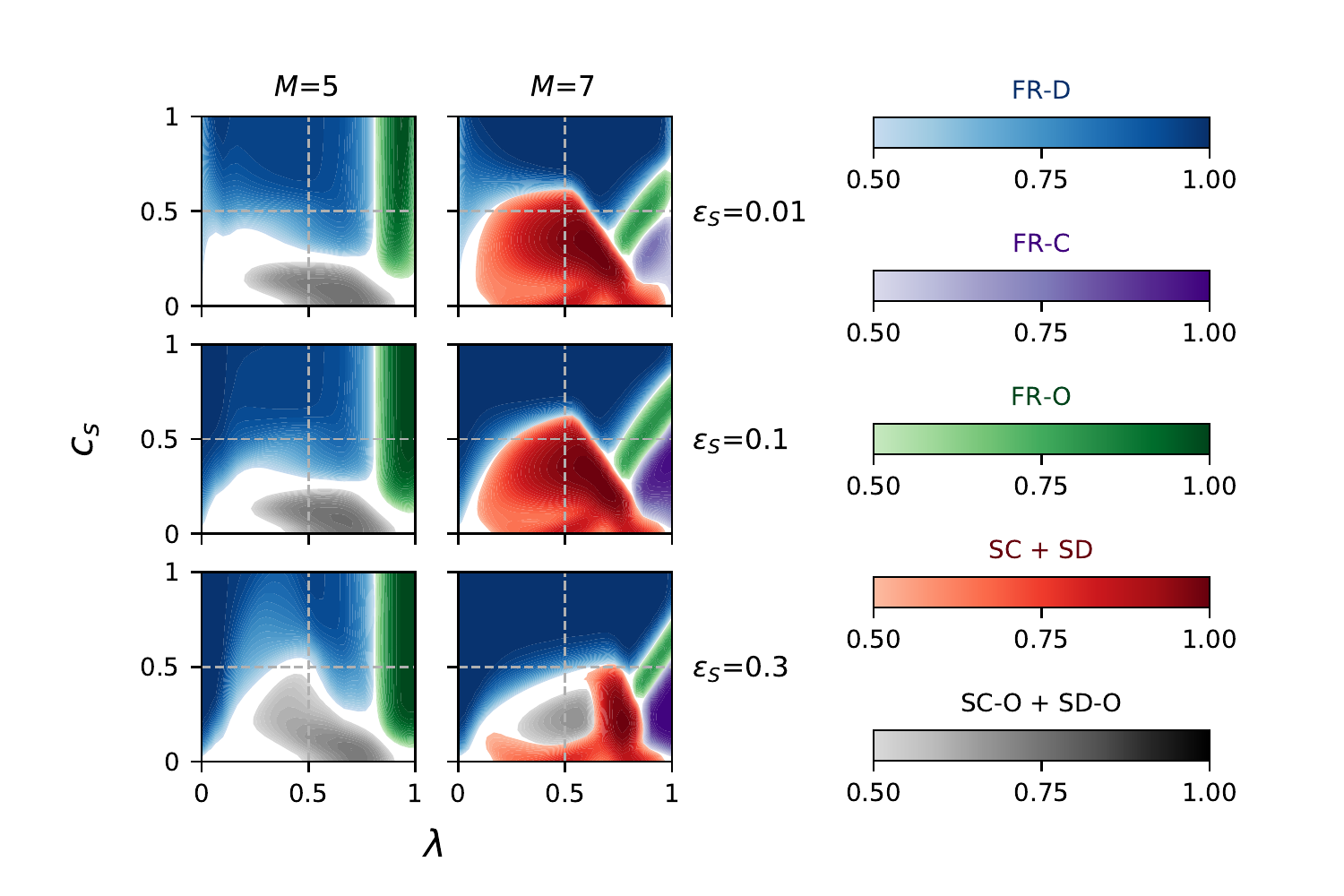}}
		\caption{\textbf{Stationary distribution of the main
                    groups of strategies for different values of
                    $\epsilon_S$}.   Model
                  parameters: $Q=N/2$, $\omega=1$, $\beta=1$, $r=10$,
                  $c=1$, $\epsilon=0.01$, $N=9$, and $Z=100$.}
		\label{fig:groupAGR_eps-w}
	\end{figure}
	
	\begin{figure}
		\centerline{\includegraphics[width=12cm]{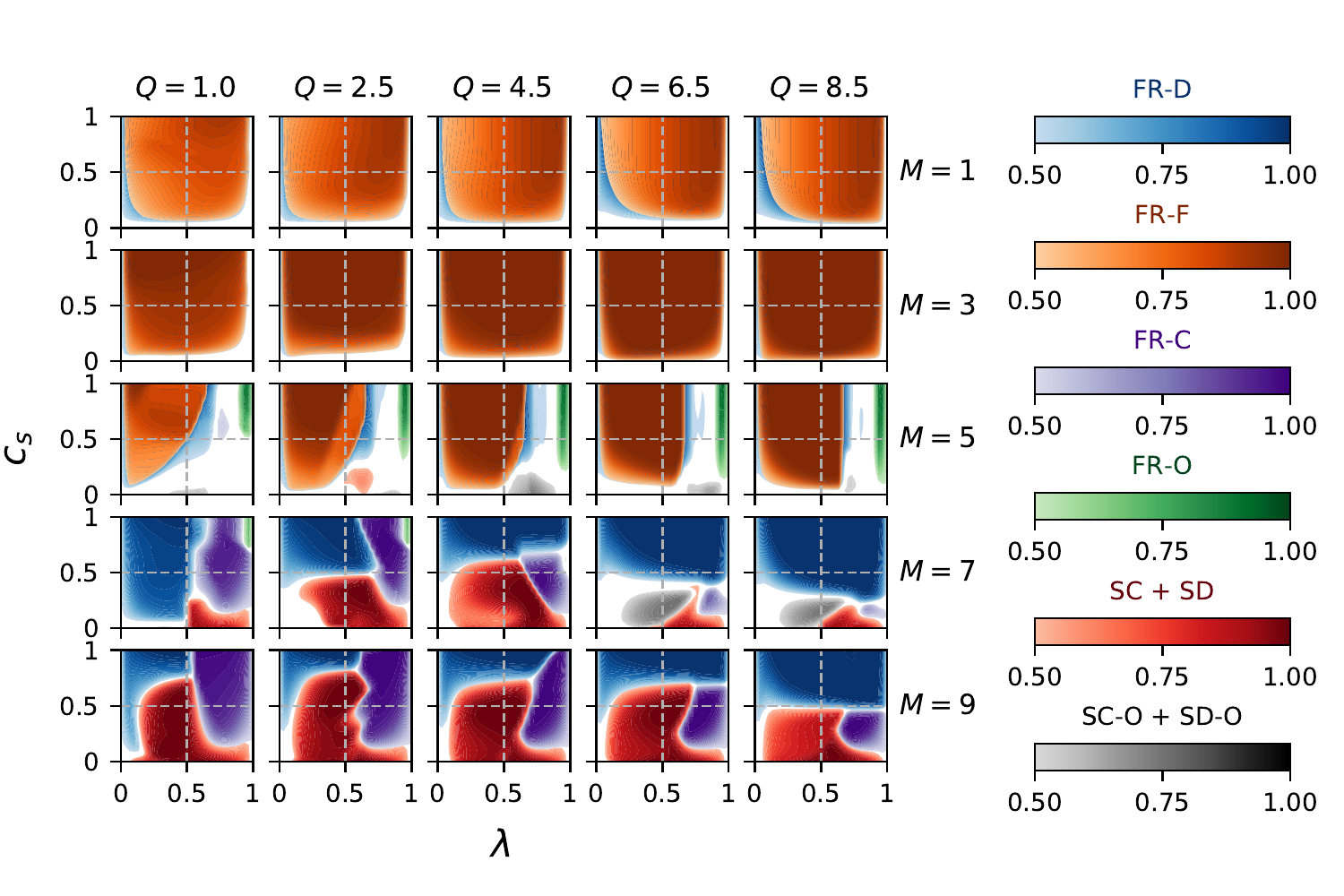}}
		\caption{\textbf{Stationary distribution of the main
                    groups of strategies  for the benefit-perceived
                    scenario under no resource}.  Model parameters: $\omega=1$, $\beta=1$,
                  $r=10$, $c=1$, $\epsilon=0.01$, $N=9$, and $Z=100$.}
		\label{fig:bper_none}
	\end{figure}
	
	\begin{figure}
		\centerline{\includegraphics[width=12cm]{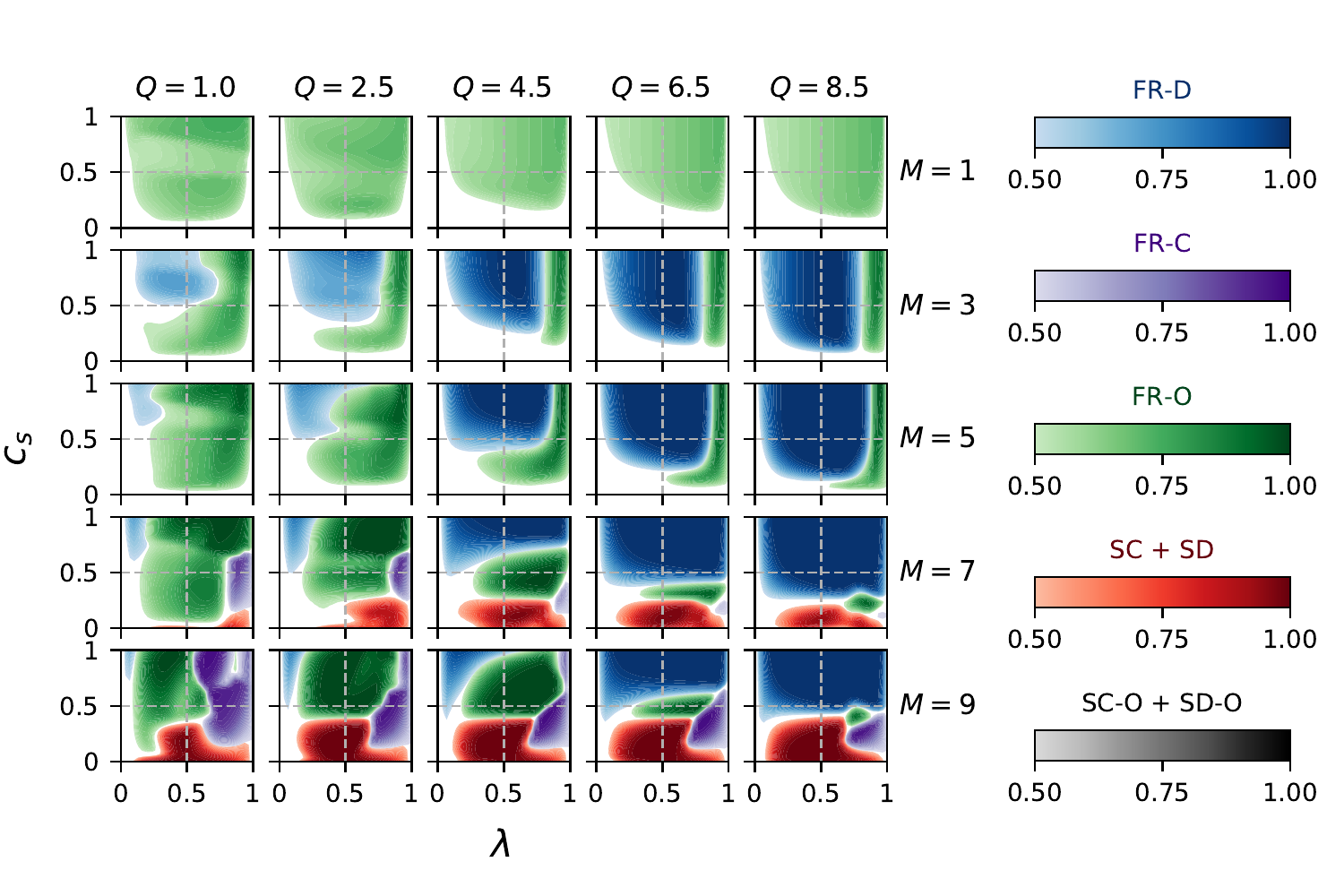}}
		\caption{\textbf{Stationary distribution of the main
                    groups of strategies for the benefit-perceived
                    scenario under abundance.} Group of strategies
                  \textbf{SC} and \textbf{SD} corresponds to
                  [10\,0$\ast$11] and[01\,110$\ast$],
                  respectively. Model
                  parameters: $\omega=1$, $\beta=1$, $r=10$, $c=1$,
                  $\epsilon=0.01$, $N=9$, and $Z=100$.}
		\label{fig:bper_abun}
	\end{figure}
	
	\begin{figure}
		\centerline{\includegraphics[width=12cm]{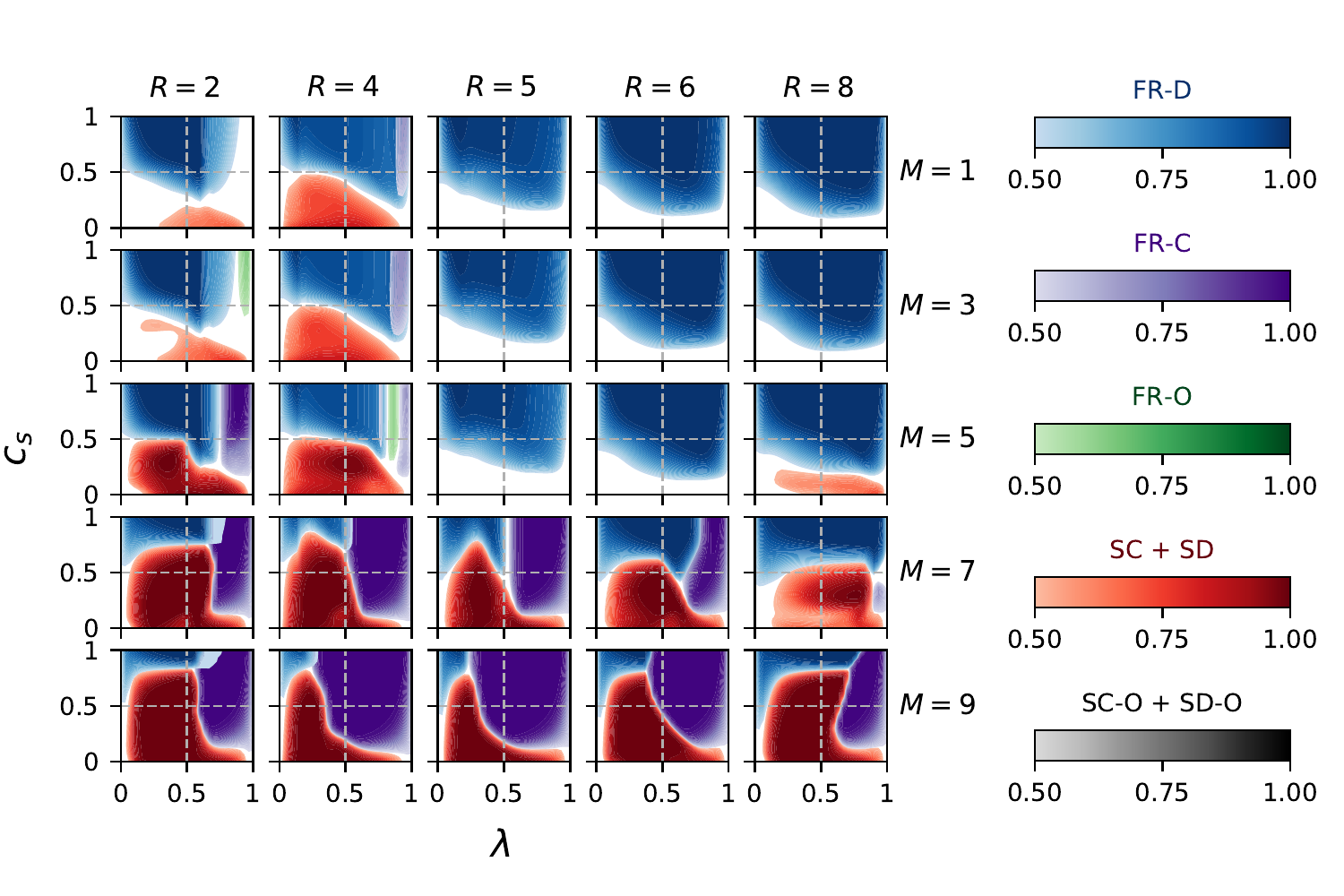}}
		\caption{\textbf{Stationary distribution of the main
                    groups of strategies for the self-aware scenario}.
                  Group of strategies change as follows:
                  \textbf{SC} corresponds to [10\,001$\ast$] when
                  $R<5$ and to [10\,$\ast$011] otherwise;
                  \textbf{SD} corresponds to [01\,1$\ast$00] when
                  $R<5$, and to [01\,11$\ast$0] otherwise; %
                  \textbf{FR-C} corresponds to
                  [00\,1$\ast$$\ast$$\ast$]
                  when $R<5$, and to
                  [00\,11$\ast$$\ast$]
                  otherwise.   Model
                  parameters: $Q=N/2$,
                  $\omega=1$,
                  $\beta=1$,
                  $r=10$, $c=1$, $\epsilon=0.01$, $N=9$, and $Z=100$.}
		\label{fig:self}
	\end{figure}